\documentclass[a4paper,11pt]{article}
\pdfoutput=1 

\usepackage{jcappub} 

\usepackage[T1]{fontenc} 
\usepackage{bm,xcolor}
\usepackage{comment}
\usepackage[nameinlink,capitalise]{cleveref}
\usepackage{multicol,multirow}

\let\Gamma\varGamma
\let\Delta\varDelta
\let\Theta\varTheta
\let\Xi\varXi
\let\Pi\varPi
\let\Sigma\varSigma
\let\Upsilon\varUpsilon
\let\Phi\varPhi
\let\Psi\varPsi
\let\Omega\varOmega

\newcommand{\de}{\mathrm{d}}
\newcommand{\fnl}{f_\mathrm{NL}}
\newcommand{\euc}{\textit{Euclid}}
\newcommand{\rom}{\textit{Roman}}
\newcommand{\hi}{H\textsc{i}}
\newcommand{\oii}{O\textsc{ii}}
\newcommand{\oiii}{O\textsc{iii}}

\newcommand{\mnras}{MNRAS}
\newcommand{\jcap}{JCAP}
\newcommand{\apj}{ApJ}
\newcommand{\apjl}{ApJL}
\newcommand{\prd}{Phys. Rev. D}
\newcommand{\aap}{A\&A}
\newcommand{\prl}{Phys. Rev. Lett}
\newcommand{\pasa}{Publications of the Astron. Soc. of Australia}
\newcommand{\physrep}{Phys. Rep.}
\newcommand{\pasp}{PASP}

\title{Radio-optical synergies at high redshift to constrain primordial non-Gaussianity}

\author[a,b]{Matilde Barberi Squarotti,}
\author[a,b,c,d]{Stefano Camera,}
\author[d,e,f]{and Roy Maartens}

\affiliation[a]{Dipartimento di Fisica, Universit\`a degli Studi di Torino, 10125 Torino, Italy}
\affiliation[b]{INFN -- Istituto Nazionale di Fisica Nucleare, Sezione di Torino, 10125 Torino, Italy}
\affiliation[c]{INAF -- Istituto Nazionale di Astrofisica, Osservatorio Astrofisico di Torino, 10025 Pino Torinese, Italy}
\affiliation[d]{Department of Physics \& Astronomy, University of the Western Cape, Cape Town 7535, South Africa}
\affiliation[e]{Institute of Cosmology \& Gravitation, University of Portsmouth, Portsmouth PO1 3FX, United Kingdom}
\affiliation[f]{National Institute for Theoretical and Computational Sciences (NITheCS), Cape Town 7535, South Africa}

\emailAdd{matilde.barberisquar@edu.unito.it}
\emailAdd{stefano.camera@unito.it}
\emailAdd{roy.maartens@gmail.com}

\abstract{We apply the multi-tracer technique to test the possibility of improved constraints on the amplitude of local primordial non-Gaussianity, $\fnl$, in the cosmic large-scale structure.  A precise measurement of $\fnl$ is difficult because the effects of non-Gaussianity mostly arise on the largest scales, which are heavily affected by  the low statistical sampling commonly referred to as cosmic variance. The multi-tracer approach suppresses cosmic variance and we implement it by  combining the information from next-generation galaxy surveys in the optical/near-infrared band and neutral hydrogen (\hi) intensity mapping surveys in the radio band. High-redshift surveys enhance the precision on $\fnl$, due to the larger available volume, and  \hi\ intensity mapping surveys can naturally reach high redshifts. In order to extend the redshift coverage of a galaxy survey, we consider different emission-line galaxy populations, focusing on the H$\alpha$ line at low redshift and on oxygen lines at higher redshift. By doing so, we cover a wide redshift range $1\lesssim z\lesssim4$. To assess the capability of our approach, we implement a synthetic-data analysis by means of Markov chain Monte Carlo sampling of the (cosmological+nuisance) parameter posterior, to evaluate the constraints on $\fnl$ obtained in different survey configurations. We find significant improvements from the multi-tracer technique: the full data set leads to a precision of $\sigma(\fnl)<1$.}

\begin{document}
\maketitle
\flushbottom

\section{Introduction}
\label{sec:intro}
The $\mathrm{\Lambda CDM}$ model can successfully explain a wide range of cosmological observations, but still leaves some issues open. In addition to  dark matter and dark energy, a phase of exponential expansion -- cosmological inflation -- is required to produce the fluctuations in the density field that seed the large-scale structure of the Universe.
A key probe of inflation is primordial non-Gaussianity \cite{2014PDU.....5...75M,2022arXiv220308128A,2004PhR...402..103B}.

The simplest inflationary models generate primordial fluctuations that follow a Gaussian distribution, but many scenarios of inflation predict departures from Gaussianity, which may be quantified by the local primordial non-Gaussianity parameter, $\fnl$ \cite{2004PhR...402..103B,2018arXiv181208197C}. The effects of non-Gaussianity can be probed in various ways, the most prominent being a measurement of a non-vanishing primordial bispectrum \cite{2020A&A...641A...9P,2011JCAP...04..006B,2011JCAP...11..038C,2013PhRvD..88h3502P,2001PhRvD..63f3002K}. However, the power spectrum  of clustering of biased tracers of the underlying matter density field also exhibits a peculiar scale-dependence on the largest scales \cite{2008PhRvD..77l3514D,2008ApJ...677L..77M}. A measurement of $\fnl \neq 0$ from the power spectrum would  allow us to rule out entire classes of inflationary models. 

Currently, the tightest constraints on $\fnl$ come from the bispectrum of the {cosmic microwave background \cite{2020A&A...641A...9P}}, but observations of the large-scale structure with next-generation cosmological surveys {will attain} competitive constraints on $\fnl$ \cite{2011JCAP...04..006B,2015PhRvD..92f3525A,2014MNRAS.442.2511F,2021JCAP...06..039J,2015ApJ...812L..22F,2020MNRAS.492.1513G,2021JCAP...11..010V,2020JCAP...12..031B,2022JCAP...01..033B,2023arXiv230209066B,2018MNRAS.478.1341K}. In the optical/near-infrared band, experiments like the European Space Agency's Euclid satellite,\footnote{\url{www.euclid-ec.org}} the Nancy Grace Roman Space Telescope\footnote{\url{https://roman.gsfc.nasa.gov}} (\rom\ hereafter), the Legacy Survey of Space and Time at the Vera C.\ Rubin Observatory,\footnote{\url{www.lsst.org}} and the Dark Energy Spectroscopic Instrument DESI,\footnote{\url{https://www.desi.lbl.gov}} will perform large galaxy surveys with high sensitivity and up to high redshifts. At much longer wavelengths, the SKA Observatory\footnote{\url{https://www.skao.int}} (SKAO) will cover a large area of the southern sky using the neutral hydrogen (\hi) intensity mapping technique to map the unresolved emission in the 21-cm line from radio galaxies residing in the large-scale structure. {Optical and radio galaxy surveys trace} two {(mainly)} independent tracers of the underlying matter distribution and they are therefore complementary. This allows us to apply the multi-tracer method \cite{2009PhRvL.102b1302S,2013MNRAS.432..318A,2009JCAP...10..007M,2019MNRAS.489.1950B} to use jointly the information from both the galaxies and the \hi\ distribution: by correlating independent biased tracers of the large-scale structure the cosmic variance error is strongly mitigated. 

The paper is organised as follows: in \cref{sec:formalism} we present the model of the power spectrum in auto- and cross-correlation, including the effects of non-Gaussianity on the power spectrum and we present the formalism of the multi-tracer technique. In \cref{sec:data}, we describe the mock data sets employed in the analysis. In \cref{sec:snr}, we evaluate the signal-to-noise ratio to asses the significance of the detection of primordial non-Gaussianity. The analysis is detailed in \cref{sec:analysis}, and the results in \cref{sec:results}. Conclusions are drawn in \cref{sec:conclusion}.

\section{Power spectrum {and multi-tracer technique}}
\label{sec:formalism}
The large-scale structure of the Universe can be studied by measuring its statistical properties. Two-point statistics have been the major target of cosmological observational campaigns, including next-generation surveys such as those mentioned above. In Fourier space, {the two-point statistic relevant for inhomogeneities in the matter distribution is} the power spectrum of the perturbations of the matter over-density. Since the matter distribution cannot be observed directly, to sample it we use (biased) tracers, such as galaxies or \hi. In the linear regime, the power spectrum of two tracers can be written as a function of the wavevector $\bm{k}$ and the redshift $z$ as 
\begin{equation} \label{pabk}
    P_{{AB}}(k,\mu,z)=\left[ b_{{A}}(z) +f(z)\,\mu^2 \right]\,\left[ b_{{B}}(z) +f(z)\,\mu^2 \right]
\,P_{\rm lin}(k,z)\;,
\end{equation}
where $A,B$ are tracer labels and $\mu$ is the cosine of the angle between the wavevector $\bm{k}$ and the line-of-sight direction (which we take as opposite to that of incoming photons). In the expression above, $b_A$ is the linear bias of tracer $A$, $f=-\de\ln D/\de\ln(1+z)$ is the growth rate, where $D(z)$ is the linear growth factor, and the term $f\,\mu^2$ describes the effects of redshift-space distortions (RSD) in the linear regime. Finally, $P_\mathrm{lin}\propto D^2(z)\,T^2(k)\,\mathcal{P}_\mathcal{R}(k)$ is the linear matter power spectrum, with $T(k)$ the transfer function (normalised such that $T=1$ for $k\to0$) and $\mathcal{P}_\mathcal{R}(k)$ the dimensionless power spectrum of the primordial curvature perturbation.

When the two tracers are the same, $A=B$, we talk about auto-correlation power spectrum, while if {$A\ne B$}, it is a cross-correlation. Moreover, analysing auto-correlations by themselves will be referred to as a single-tracer analysis, whereas considering both auto- and cross-correlations at the same time leads to a multi-tracer analysis.

The effect of primordial non-Gaussianity of the local type is to introduce a scale dependence in the linear bias \cite{2004PhR...402..103B,2008PhRvD..78l3519M,2008PhRvD..77l3514D,2018arXiv181208197C,2020JCAP...12..031B,2021JCAP...05..015M,2022JCAP...01..033B,2023JCAP...01..023L}: 
\begin{equation} \label{Delta_b}
    b_A(z)\to \tilde b_A(z,k)=  b_A(z)
    +\fnl\,b_{A\phi}(z) \,\frac{3\,\Omega_{\rm m,0}}{2\,c^2\,D(z)}\,\frac{H_0^2}{T(k)\,k^2}\;.
\end{equation}
Here, $b_{A\phi}$ is the primordial non-Gaussian bias factor, where $\phi$ is the potential deep in the matter era. This  bias, like the Guassian bias $b_A$, depends on the properties of tracer $A$ -- but the relation between $b_A$ and $b_{A\phi}$ is still poorly understood. Simulations indicate that this relation is sensitive to the assembly
history and the selection criteria of the tracer \cite{2020JCAP...12..031B,2021JCAP...05..015M,2022JCAP...01..033B,2023JCAP...01..023L}. Since $b_{A\phi}$
is completely degenerate with $\fnl$, we cannot constrain $\fnl$ without a theoretical model for $b_{A\phi}$. Typically, a strong theoretical prior has been imposed by assuming universality of the halo mass function. In this case,
\begin{align}\label{unirel}
b_{A\phi} = 2\,\delta_{\rm c}\,\left(b_A-1\right)\;,   
\end{align}
where $\delta_{\rm c}\simeq1.686$ is the critical density contrast for collapse in the linear regime.

Nearly all observational constraints from galaxy surveys on $\fnl$ to date rely on this universality assumption -- with a few exceptions  (e.g. \cite{2022PhRvD.106d3506C,2022JCAP...11..013B}). Similarly, most forecast constraints are based on assuming the universality relation. Further progress in uncovering the relation between  $b_A$ and $b_{A\phi}$ is needed in order to break the degeneracy without invoking strong theoretical priors. 
Clearly the best-fit values of  $\fnl$ will be biased by using incorrect models of $b_{A\phi}$, and $\fnl$ errors will be increased by  uncertainties on $b_{A\phi}$ in a given model.
However, we highlight the point that a multi-tracer analysis is significantly more robust to these problems than a single-tracer analysis \cite{2023arXiv230209066B}.
Here we assume that the degeneracy is broken and that the constraints on $\fnl$ are not very sensitive to the detailed form of the relation between $b_{A\phi}$ and $b_A$. Then we use \cref{unirel} as the simplest proxy for this relation.

The bias correction for non-Gaussianity is proportional to $T(k)^{-1}k^{-2}$, which implies that the effects become manifest at ultra-large scales, $T\to 1$, 
where the cosmic variance is larger due to the decrease of observable independent modes. To overcome this limit, we apply the multi-tracer technique, which relies on the fact that the correlation of independent biased tracers of the same underlying matter distribution allows us to eliminate the error of the cosmic variance \cite{2009PhRvL.102b1302S,2009JCAP...10..007M,2013MNRAS.432..318A}. This technique consists in analysing simultaneously the auto- and cross-correlation power spectra of two tracers in a single, joint data vector. Namely, in our case,
\begin{equation}
    {\bm P} = \big\{ P_{\rm g\,g}, P_{\rm g\,\hi}, P_{\rm\hi\,\hi} \big\} \; .
\end{equation}

The full covariance matrix associated to the multi-tracer power spectra is given by \cite{2020JCAP...12..031B,2023arXiv230504028K}
\begin{equation}
    \mathsf{Cov}({\bm P},{\bm P})=\frac{2}{N_{\bm k}}
    \begin{bmatrix}
        \tilde P_{\rm g\,g}^2     & \tilde P_{\rm g\,g}\,\tilde P_{\rm g\,\hi}     & \tilde P_{\rm g\,\hi}^2 \\
        \tilde P_{\rm g\,g}\,\tilde P_{\rm g\,\hi}     &  \dfrac{1}{2}\,\left( \tilde P_{\rm g\,g}\,\tilde P_{\rm\hi\,\hi} + \tilde P_{\rm g\,\hi}^2 \right)   & \tilde P_{\rm\hi\,\hi}\,\tilde P_{\rm g\,\hi} \\
        \tilde P_{\rm g\,\hi}^2     & \tilde P_{\rm\hi\,\hi}\,\tilde P_{\rm g\,\hi}     & \tilde P_{\rm\hi\,\hi}^2 \; ,
    \end{bmatrix}
    \label{eq:covariance}
\end{equation}
where 
$\tilde P_{AB}=P_{AB}+P^{\mathrm{noise}}_{AB}\,\delta_{AB}^{\rm K}$, $P^{\mathrm{noise}}_{AB}$ is the (statistical or instrumental) noise, and $\delta_{AB}^{\rm K}$ is the Kronecker delta. The term $N_{\bm k}(z)$ represents the number of independent modes within an observed volume $V(z)$ at given $z$ and $\bm k$, i.e.\
\begin{equation}
    N_{\bm k}(z)=\frac{V(z)}{(2\,\pi)^3}\,2\,\pi\, k^2\,\Delta k(z)\,\Delta\mu\;,
\end{equation}
with $\Delta k(z)$ and $\Delta \mu$ the width of the bins in $k=|\bm k|$ and $\mu$, respectively. (Note that, in principle, the $k$-binning may depend on redshift.) For a survey covering a sky fraction $f_{\mathrm{sky}}$, the comoving volume observed at redshift $z$ is
\begin{equation}
    V(z)=\frac{4\,\pi\,f_{\rm sky}}3\left[\chi^3\!\left(z+\frac{\Delta z}2\right)-\chi^3\!\left(z-\frac{\Delta z}2\right)\right]\;,
\end{equation}
where $\chi(z)$ is the comoving distance up to redshift $z$.

On the ultra-large scales where the signal of local primordial non-Gaussianity is strongest, general relativistic `projection' effects -- i.e.\ effects from observing on the past lightcone -- can also become non-negligible \cite{2010PhRvD..82h3508Y,2011PhRvD..84f3505B,2011PhRvD..84d3516C,2012PhRvD..85d1301B,2012PhRvD..85b3504J,2012PhRvD..86f3514Y}. The observed overdensity of tracer $A$ is then of the form
\begin{align}
 \delta_A =  \tilde b_A\,\delta_{\rm m} -\frac{(1+z)}{H}\,\bm{n} \cdot \bm{\nabla}(\bm{n}\cdot\bm{v})+ \left(5\,s_A-2 \right)\kappa  + \delta^{\rm dp}_A\;,
\end{align}
where $\bm{n}$ is the line-of-sight direction, $\bm v$ is the peculiar velocity of matter (assumed equal to that of tracer $A$). The second term is the standard redshift-space distortion of overdensity. The third term is the lensing distortion of overdensity, where $s_A$ is the magnification bias of tracer $A$ and $\kappa$ is the lensing convergence. The last term, $\delta^{\rm dp}_A$, contains Doppler and potential distortions of overdensity (including Sachs-Wolfe, integrated Sachs-Wolfe and time delay terms). A natural way to include all these effects in the 2-point correlations is via the angular power spectra (which also naturally include the full wide-angle effects). This allows for a theoretically correct analysis of local primordial non-Gaussianity in clustering, as performed in \cite{2015MNRAS.448.1035C} and then generalised by e.g. \cite{2015ApJ...814..145A,2015PhRvD..92f3525A,2015ApJ...812L..22F}.

The question is: what is the consequence for $\fnl$ of neglecting the general relativistic projection effects? This was first addressed in the case of lensing convergence in photometric surveys by \cite{2011PhRvD..83l3514N} (see e.g. \cite{2022A&A...662A..93E} for recent work on this case). Then all general relativistic effects were included (in spectroscopic surveys) by \cite{2015MNRAS.451L..80C} (for recent work, see e.g. \cite{2021JCAP...12..004V}, which also uses a multi-tracer analysis, and \cite{2023PhRvL.131k1201F}). It turns out that the estimate of the best-fitting value of $\fnl$ can be strongly biased by the neglect of relativistic projection effects -- but the error on $\fnl$ is not significantly affected \cite{2021JCAP...12..004V}. Since we are performing forecasts of $\sigma(\fnl)$ and not measurements of $\fnl$, it is reasonable for us to neglect the relativistic projection effects and use the Fourier power spectra given by \cref{pabk}.

\section{Data sets}
\label{sec:data}
The first step for building a synthetic data set both for galaxies and \hi\ is defining the observed cosmic volume, i.e.\ the depth and the sky coverage of the reference experiments. The observed volume, and therefore the redshift, define in turn the scales that can be probed by the surveys. The relation between the redshift and the observed scale arises from the fact that the observed sky volume is a redshift dependent quantity. Assuming that we are observing a sky cube of volume $V$, the minimum wavenumber, $k_{\mathrm{min}}$, corresponding to the largest scale observed, can be expressed as
\begin{equation}
    k_{\mathrm{min}}(z)=\frac{2\,\pi}{V^{1/3}(z)}\;. \label{kmin}
\end{equation}

The interval in $k$ is then different for each observed redshift and it is defined by this minimum value, decreasing with the redshift, while the upper limit is set to be the largest $k$ before non-linearities need to be taken into account \cite{2003MNRAS.341.1311S,2020JCAP...03..065M,2021JCAP...06..039J}
\begin{equation}
    k_{\mathrm{max}}(z)=0.08\,(1+z)^{2/(2+n_{\rm s})} \, h \, {\rm Mpc^{-1}} \;, \label{kmax}
\end{equation}
where \(n_{\rm s}\) is the slope of the primordial curvature power spectrum. This choice, resulting in $0.085<k_{\rm max}\,{\rm Mpc}<0.16$, is made in order to avoid  non-linear scales where theoretical understanding is still poor. Moreover, since the primordial non-Gaussianity signal is strongest on the largest scales, we prefer to be conservative and not include even mildly non-linear scales.

In our analysis, we adopt 12 bins in the range $z\in[0.85,4.0]$, 10 linear bins in $k$, and 10 equi-spaced bins in $\mu$ spanning all the possible values between $\mu=-1$ and $\mu=1$. The redshift bins have a variable width $\Delta z \sim 0.2$, the width being chosen to follow the transition between galaxy types \cite{2020MNRAS.495.1340F}, and the same division was adopted for the \hi\ sample in order to match the binning between radio and optical tracers. Finally, the binning in $k$ is done using 10 bins in order to ensure that the condition $\Delta k \geq k_{\rm min}$ is satisfied.

\begin{figure}[tbp]
\centering 
\includegraphics[width=.51\textwidth]{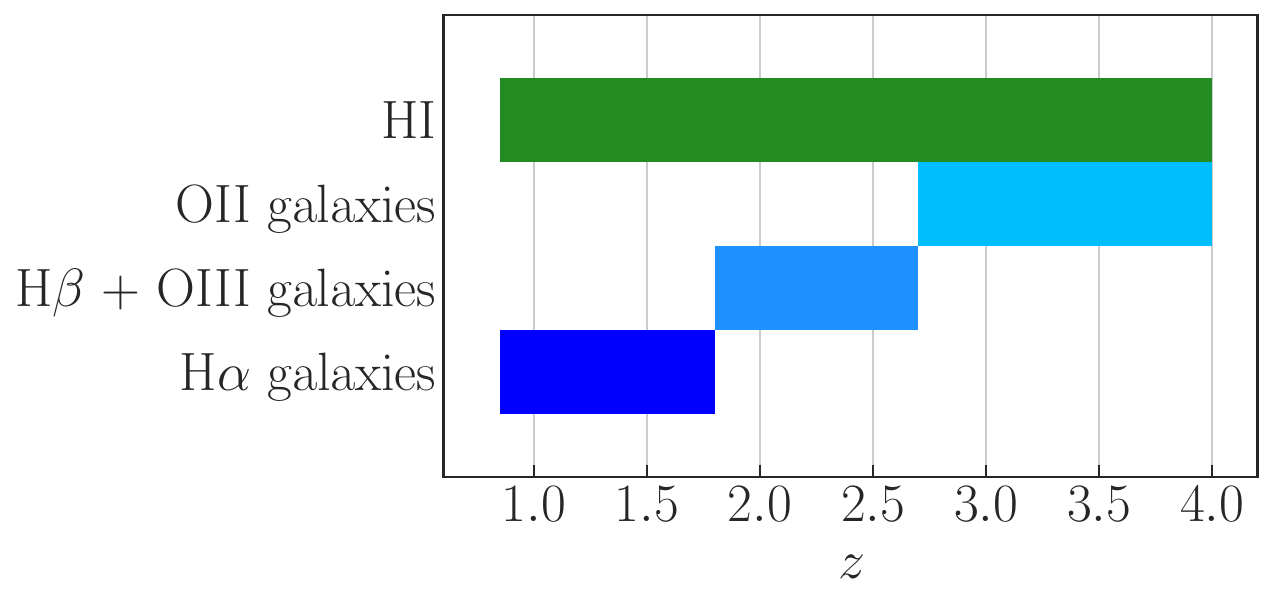}
\hfill
\includegraphics[width=.48\textwidth]{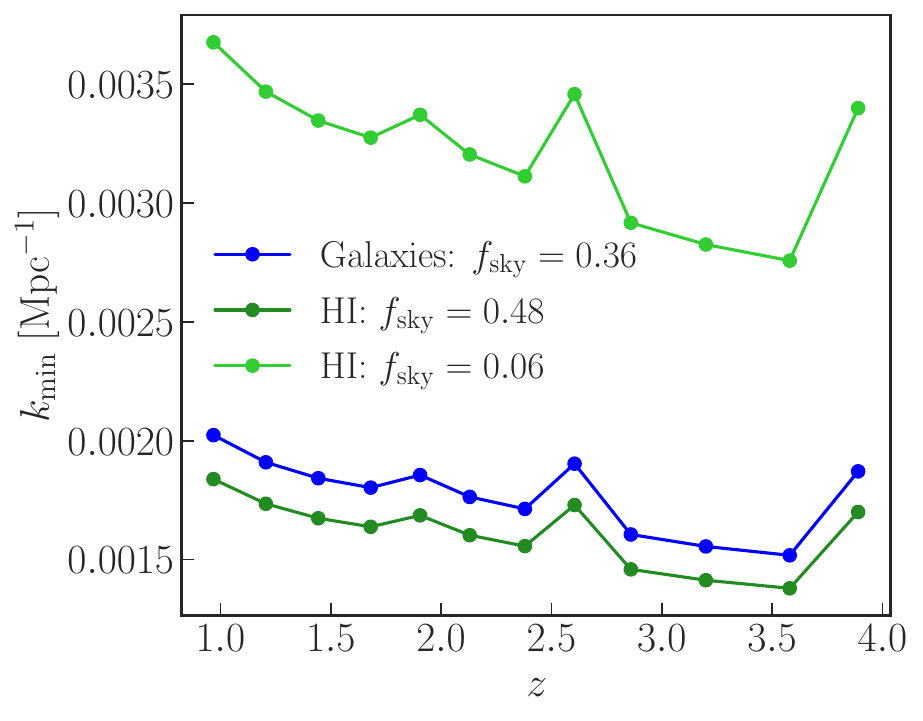}
\caption{\label{redshiftandscale} Redshift coverage of the surveys (left): the \hi\ intensity mapping survey  spans the whole interval from $z=0.85$ to $4.0$; conversely for a spectroscopic galaxy survey different galaxy populations are needed. Largest scale observed, i.e.\  smallest wavenumber $k_{\rm min}$ (right): as a function of  redshift for different sky fractions: $f_{\rm sky}=0.36$ (reference galaxy survey), $f_{\rm sky}=0.48$ (\hi\ survey), and $f_{\rm sky}=0.06$ (\hi\ survey with foreground avoidance using $N_{\rm fg}=2$) (see discussion in \cref{HIdata}). At higher redshifts, larger scales can be accessed, however at the redshifts of transition between galaxy types there is a small spike since the redshift bin is narrower.}
\end{figure}

\subsection{Galaxy samples}
\label{ssec:galdata}
As a benchmark, we consider an emission-line galaxy survey. We assume a target sample at low redshift of H$\alpha$ galaxies between $z\simeq0.9$ and $z\simeq1.7$. At higher redshift, oxygen lines, i.e.\ the \oiii\ line and the \oii\ doublet, will be used to identify galaxies, considering also contributions from the overlapping H$\beta$ line. This is in line with the capabilities of present and near-future cosmological experiments like \euc\ and \rom. Note, however, that for the former, oxygen-line galaxies will in fact be interlopers, which must be identified correctly to reach the completeness and purity required for the target sample. But once they have been identified, they can be used as additional samples of galaxies to extend the redshift range [\citealp{2020MNRAS.495.1340F}; see also \citealp{2021MNRAS.505.2784Z,2023MNRAS.523.2498M}]. Specifically, we consider a sample of H$\beta+$\oiii\ galaxies between $z\simeq1.9$ and $z\simeq2.6$ and a sample of \oii\ galaxies in the range $2.8\lesssim z\lesssim4$: the transition between different galaxy types is represented in the left panel of \cref{redshiftandscale}. For the sky coverage, we adopt $15\,000\,\deg^2$. The impact of a different sky coverage is addressed in \cref{appendix:apa}.

\begin{figure}[tbp]
\centering 
\includegraphics[width=.65\textwidth]{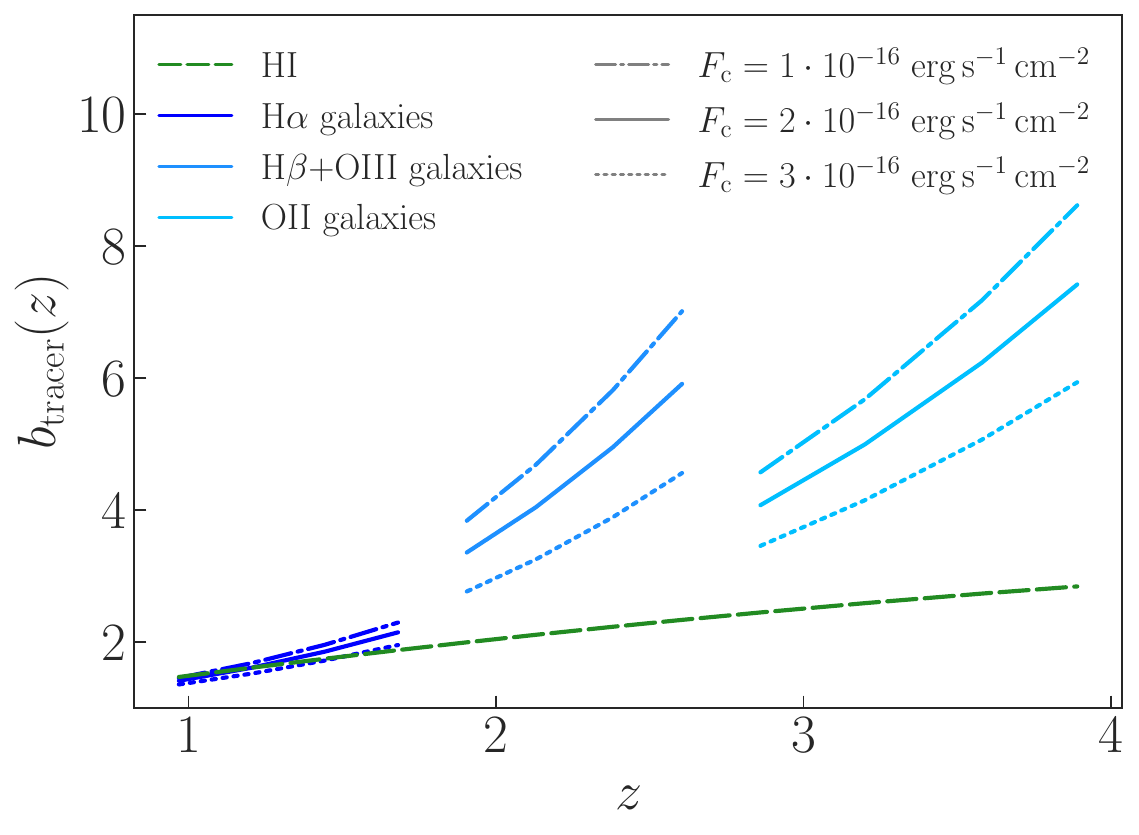}
\caption{\label{bias}
Linear bias of \hi\ and of different galaxy types. The galaxy bias is computed varying the flux limit $F_{\rm c}$: dotted and dash-dotted lines represent respectively higher and lower flux limits compared to the reference value $F_{\rm c} = 2 \times 10^{-16} \; {\rm erg \, s^{-1} \, cm^{-2}}$ (solid). H$\alpha$-line galaxies are characterised by a bias closer to unity and do not present a relevant dependence on the flux limit. All other galaxy types considered in our data set at higher redshift have a higher bias and a stronger dependence both on the redshift and on the flux limit of the survey. The green dashed line corresponds to the \hi\ bias parametrised as in \cref{hibias}.}
\end{figure}

\begin{table}
\centering
\begin{tabular}{|c|c|c|c|c|c|}
\hline
& $a$               & $b$               & $c$                & $d$               & $e$               \\
\hline
H$\alpha$ & $0.844 \pm 0.031$ & $0.116 \pm 0.02$  & $42.623 \pm 0.132$ & $1.186 \pm 0.387$ & $1.756 \pm 0.106$ \\
\hline
H$\beta$+\oiii  & $0.837 \pm 0.036$ & $0.136 \pm 0.027$ & $42.2 \pm 0.119$   & $1.409 \pm 0.395$ & $1.681 \pm 0.117$ \\
\hline
\oii       & $0.816 \pm 0.03$  & $0.118 \pm 0.022$ & $42.993 \pm 0.407$ & $1.38 \pm 0.636$  & $1.855 \pm 0.109$ \\
\hline
\end{tabular}
\caption{\label{tablebias}Values of the parameters of the fitting formula used for the galaxy bias.}
\end{table}
Regarding the properties of the various galaxy samples, we parametrise their (linear) bias as a function of the galaxy type, the redshift, and the flux limit of the survey. We adopted the 5-parameter model described in \cite{2020MNRAS.493..747P}, which provides the fitting formula
\begin{equation}
    b_{\rm g}(x,z)=a+b\,(1+z)^e\,\left[ 1+e^{(x-c)\,d} \right]\;.
\end{equation}
Here $x$ is related to the flux limit of the survey through $x=\log_{10}L_{\rm min} =\log_{10}\left[ 4\,\pi\,D_{\rm L}^2(z)\,F_{\rm c} \right]$, where $D_{\rm L}$ is the luminosity distance and cgs units are used. The five parameters are set according to the galaxy type: their values are presented in \cref{tablebias}. The dependence of the galaxy bias on the redshift is displayed in \cref{bias} for different galaxy types and flux limits. We also investigate other possible parametrisations in \cref{appendix:apa}.

The expected number density of galaxies per unit volume in each redshift bin can be calculated by integrating the galaxy luminosity function $\Phi(L,z)$, i.e.\
\begin{equation} \label{galng}
    {\bar n}(z)= \int_{L_{\rm min}/L_{\ast}}^{L_{\rm max}/L_{\ast}} {\rm d}\left(\frac{L}{L_{\ast}}\right) \Phi(L,z) \; ,
\end{equation}
where the minimum luminosity is determined by the flux limit of the survey, while the maximum luminosity can be safely set to infinity. The functional form of $\Phi(L,z)$ is chosen in order to capture the differences between galaxy types, following \cite{2020MNRAS.495.1340F}. For the H$\alpha$ sample we adopted the broken power law luminosity function described in \cite{2016A&A...590A...3P},
\begin{equation}
\Phi(L,z)=\Phi_{\ast}\left(\frac{L}{L_{\ast}}\right)^{\alpha} \left[ 1+({\rm e}-1) \left(\frac{L}{L_{\ast}}\right)^{\nu} \right]^{-1} \; ,
\end{equation}
where $\Phi_{\ast}$ is fixed to its value at $z=0$ while the reference luminosity $L_{\ast}$ is given by
\begin{equation}
    \log_{10}L_{\ast}(z)=\log_{10}L_{\ast}(z=\infty)+\left(\frac{1.5}{1+z}\right)^{\beta} \log_{10}\frac{L_{\ast}(z=0.5)}{L_{\ast}(z=\infty)} \; .
\end{equation}
The values of the parameters that characterise the H$\alpha$ luminosity function are $\alpha=-1.587$, $\nu=2.288$, $\beta=1.615$, $\Phi_{\ast}(z=0)=10^{-2.920} \; {\rm Mpc^{-3}}$, $L_{\ast}(z=0.5)=10^{41.733} \; {\rm erg \, s^{-1}}$, $L_{\ast}(z=\infty)=10^{42.956} \; {\rm erg \, s^{-1}}$.  
For the H$\beta$+\oiii\ and the \oii\ samples, we used a Schechter luminosity function,
\begin{equation}
\Phi(L,z)=\Phi_{\ast}\left(\frac{L}{L_{\ast}}\right)^{\alpha} {\rm e}^{-L/L_{\ast}} \; ,
\end{equation}
where the values of  $\alpha$, $\Phi_{\ast}(z_i)$ and $ L_{\ast}(z_i)$ are given in \cref{tableflum} for each galaxy type, according to the estimates of \cite{2015MNRAS.452.3948K}.

Further comments and analysis on the observed number density of interlopers are presented in \cref{appendix:apa}.

\begin{table}
\centering
\begin{tabular}{ccccccccc} 
\hline
\multicolumn{4}{c}{H$\beta$+\oiii}                                                        && \multicolumn{4}{c}{\oii}                                                                 \\ 
\hline
$z$    & $\log_{10}(\Phi_{\ast})$ & $\log_{10}(L_{\ast})$   & $\alpha$                 && $z$    & $\log_{10}(\Phi_{\ast})$ & $\log_{10}(L_{\ast})$   & $\alpha$                  \\ 
\hline
$0.84$ & $-2.55^{+0.04}_{-0.03}$  & $41.79^{+0.03}_{-0.05}$ & \multirow{4}{*}{$-1.60$} && $1.47$ & $-2.25 \pm 0.04$         & $41.86 \pm 0.03$         & \multirow{4}{*}{$-1.30$}  \\ 
\cline{1-3}\cline{6-8}
$1.42$ & $-2.61^{+0.10}_{-0.09}$  & $42.06^{+0.06}_{-0.05}$ &                          && $2.25$ & $-2.48^{+0.08}_{-0.09}$  & $42.34^{+0.04}_{-0.03}$  &                           \\ 
\cline{1-3}\cline{6-8}
$2.23$ & $-3.03^{+0.21}_{-0.26}$  & $42.66 \pm 0.13$        &                          && $3.34$ & $-3.07^{+0.63}_{-0.70}$  & $42.69^{+0.31}_{-0.23}$  &                           \\ 
\cline{1-3}\cline{6-8}
$3.24$ & $-3.31^{+0.09}_{-0.26}$  & $42.83^{+0.19}_{-0.17}$ &                          && $4.69$ & $-3.69^{+0.33}_{-0.28}$  & $42.93^{+0.018}_{-0.24}$ &                           \\
\hline
\end{tabular}
\caption{\label{tableflum}Values of the parameters that characterise the luminosity function of the H$\beta$+\oiii\ and the \oii\ galaxy samples. Units are ${\rm Mpc^{-3}}$ for the  number densities $\Phi_{\ast}$ and ${\rm erg \, s^{-1}}$ for the luminosities $L_{\ast}$.}
\end{table}
For a galaxy survey, the noise in the power spectrum corresponds to a shot-noise term given by the inverse of the observed galaxy number density in a given redshift bin, namely
\begin{equation}
    P^{\mathrm{noise}}_{\rm g\,g}(\bar z_i)=\frac{1}{\bar n_i}\;,
\end{equation}
where $\bar z_i$ is the mean redshift of the $i$th redshift bin and $\bar n_i$ is the number density of galaxies per unit volume in that bin. The impact of shot noise is the more severe the higher the redshift, as a consequence of the flux limit of a spectroscopic survey: galaxies at high $z$ are fainter and difficult to detect, and this results in a lower observed number density. We consider a flux limit of $F_{\rm c} = 2 \times 10^{-16} \; {\rm erg \, s^{-1} \, cm^{-2}}$ as reference and use the observed number galaxies calculated as in \cite{2020MNRAS.495.1340F}. Then we also let the flux limit vary from $F_{\rm c} = 10^{-16} \; {\rm erg \, s^{-1} \, cm^{-2}}$ to $F_{\rm c} = 3 \times 10^{-16} \; {\rm erg \, s^{-1} \, cm^{-2}}$ to evaluate the impact of the observed galaxy number density on the forecast. Note that we choose the aforementioned values to include the capabilities of present and upcoming surveys, such as \textit{Euclid} \citep{2011arXiv1110.3193L} and \textit{Roman} \citep{2022ApJ...928....1W}.

\subsection{\hi\ intensity mapping}
\label{HIdata}
The intensity mapping survey is considered to have the properties of the corresponding SKAO survey in the mid band proposed in the SKA Cosmology Red Book 2018 \cite{2020PASA...37....7S}. Such a survey will cover a sky fraction $f_{\mathrm{sky}}=0.48$ in single-dish mode, covering the whole redshift range considered for the analysis.  We model the \hi\ bias according to the following parametrisation \cite{2018ApJ...866..135V},
\begin{equation} \label{hibias}
    b_{\rm\hi}(z) = 0.842\,\left(1+0.823\,z-0.0546\,z^2\right) \;.
\end{equation}

The dominant noise term in the case of the intensity mapping is due to thermal noise of the instrument, which we model as
\begin{equation}
    P^{\mathrm{noise}}_{\rm\hi\,\hi}(z)= \frac{2\,\pi\,f_{\rm sky}\,c}{\nu _{\rm\hi}\,t_{\rm tot}\,N_{\rm d}}\,\frac{\chi^2(z)}{H(z)} \, \left[ \frac{T_{\rm sys}(z)}{\overline{T}_{\rm\hi}} \right]^2\;.
\end{equation}
In the expression above, $t_{\rm tot}$ is the total observing time, $N_{\rm d}$ is the number of dishes, while the \hi\ temperature and the system temperature are modelled respectively as
\begin{align}
    \overline{T}_{\rm\hi}(z) \, [{\rm mK}]&=0.056+0.23\,z-0.024\,z^2\;,\\
    T_{\rm sys}(z) \, [{\rm K}]&= T_{\rm d}(z)+T_{\rm sky}(z) = T_{\rm d}(z)+2.7+25\,\left[ \frac{400\,{\rm MHz}}{\nu _{\rm\hi}} (1+z) \right]^{2.75}\;.\label{thnoise}
\end{align}
For $\overline{T}_{\rm\hi}$, we use the parametrisation given in \cite{2016mks..confE..32S,2020A&A...641A...9P,2019JCAP...12..028F}, which captures the dependence of the average \hi\ brightness temperature on the comoving \hi\ density fraction, $\Omega_{\rm \hi}$, whose full expression is given in Eq.\ (2.1) of \cite{2015aska.confE..19S}. For $T_{\rm d}$, which is the dish receiver temperature, we instead follow \cite{2021JCAP...06..039J}. It is worth noticing that there is a difference of orders of magnitude between $\overline{T}_{\rm\hi}$ and $T_{\rm sys}$, meaning that the cosmological \hi\ signal is not the dominant contribution. Note also that the thermal noise is scale-independent and increases with redshift. 

In principle, a shot noise term should be present as well, as the continuous \hi\ emission is fundamentally a collection of the emission of all unresolved \hi-line galaxies. It can be derived in a halo-model framework as \cite{2018ApJ...866..135V,2017MNRAS.471.1788C,2015ApJ...812L..22F}
\begin{equation} \label{hishot}
    P_{\rm\hi\,\hi}^{\rm shot}=\frac{1}{\overline{\rho}_{\rm\hi}^2(z)}\int{\rm d}M\, n_{\rm h}(M,z)\,M_{\rm\hi}^2(M,z) \; ,
\end{equation}
where $\overline{\rho}_{\rm\hi}$ is the average comoving \hi\ density, $n_{\rm h}$ is the halo mass function, and $M_{\rm\hi}$ the average mass of neutral hydrogen within a halo of mass $M$. However, on linear scales, and especially at high redshift, the amplitude of shot noise is much lower that that of thermal noise. Since this is the regime we are mainly interested in, we neglect the shot noise term in our analysis.

Intensity mapping observations also suffer from a loss of signal at small scales in the perpendicular direction due to the low angular resolution. In an idealistic scenario, this can be modelled with a Gaussian beam perpendicular to the line-of-sight direction. It results into a damping factor affecting perpendicular modes $k_{\perp}=\sqrt{(1-\mu ^2)}\,k$ as \cite{2021JCAP...06..039J,2023MNRAS.518.6262C,2023MNRAS.523.2453C,2019MNRAS.489..153B}
\begin{equation}
    \mathcal{D}_{\rm b}(k,z,\mu)=\exp \left[ {- \frac{k_{\perp}^2\,\chi^2(z)\,\theta^2_{\rm b}(z)}{16\,\ln 2}} \right] \; ,
\end{equation}
where the size of the beam $\theta_{\rm b}$ of a dish of diameter $D_{\rm d}$ is given, in radians, by \cite{2015ApJ...803...21B}
\begin{equation}
    \theta_{\rm b}(z) = 1.22\,\frac{\lambda_{\rm\hi}(1+z)}{D_{\rm d}} \;.
\end{equation}

In the case of the intensity mapping the foreground contamination must be taken into account \cite{2014MNRAS.441.3271W,2019MNRAS.488.5452C,2020MNRAS.499.4054C,2022MNRAS.509.2048S}. Foregrounds mostly affect the largest radial scales, where they cannot be separated from the cosmological signal. We decide to treat this effect with a foreground avoidance approach \cite{2017MNRAS.471.1788C,2022MNRAS.512.2408C,2021MNRAS.502.2549S,2023EPJC...83..320J}, considering two different methods. The first one consists in applying an exponential suppression factor to the power spectrum in order to remove the radial modes $k_{\parallel}=\mu\,k$ smaller than a critical scale $k_{\parallel {\rm fg}}$. This can be modelled as
\begin{equation} \label{fgdamping}
    \mathcal{D}_{\rm fg}(k,\mu)=1- \exp \left( - \frac{k_{\parallel}^2}{k_{\parallel {\rm fg}}^2} \right) \; .
\end{equation}
The second approach is more conservative, i.e.\  all the scales corresponding  to \citep{2021MNRAS.502.2549S}
\begin{equation}\label{knfg}
 k<N_{\mathrm{fg}}\,k_{\mathrm{min}}   
\end{equation}
are excluded from the analysis, where $k_{\mathrm{min}}$ is the minimum  wavenumber in the given redshift bin and $N_{\rm fg}$ is a factor to be chosen. As a consequence, the survey's effective volume and the sky fraction must be rescaled accordingly by a factor $N_{\rm fg}^3$. Note that this method not only limits the radial $k_{\parallel}$, but also the transverse $k_{\perp}$.

Both methods result in a loss of signal at large scales, which is problematic for constraining parameters such as $\fnl$.  In the following we will consider the exponential suppression as reference for the analysis, and then we will also compare it to the results that can be obtained from the more drastic approach. The resulting \hi\ power spectrum, once the damping factors are taken into account, reads
\begin{equation}
    P_{\rm\hi\,\hi}(k,z,\mu) \rightarrow  \mathcal{D}_{\rm fg}(k,\mu) \, \mathcal{D}_{\rm b}^2(k,z,\mu) \,P_{\rm\hi\,\hi}(k,z,\mu) \; , 
\end{equation}
or, in the conservative approach of \cref{knfg},
\begin{equation}
    P_{\rm\hi\,\hi}(k,z,\mu) \rightarrow 
    \Theta(k-N_{\mathrm{fg}}\,k_{\mathrm{min}})\,
    \mathcal{D}_{\rm b}^2(k,z,\mu) \, P_{\rm\hi\,\hi}(k,z,\mu) \; ,
\end{equation}
where
$\Theta$ is the Heaviside step function.

\begin{figure}[tbp]
\centering 
\includegraphics[width=.75\textwidth]{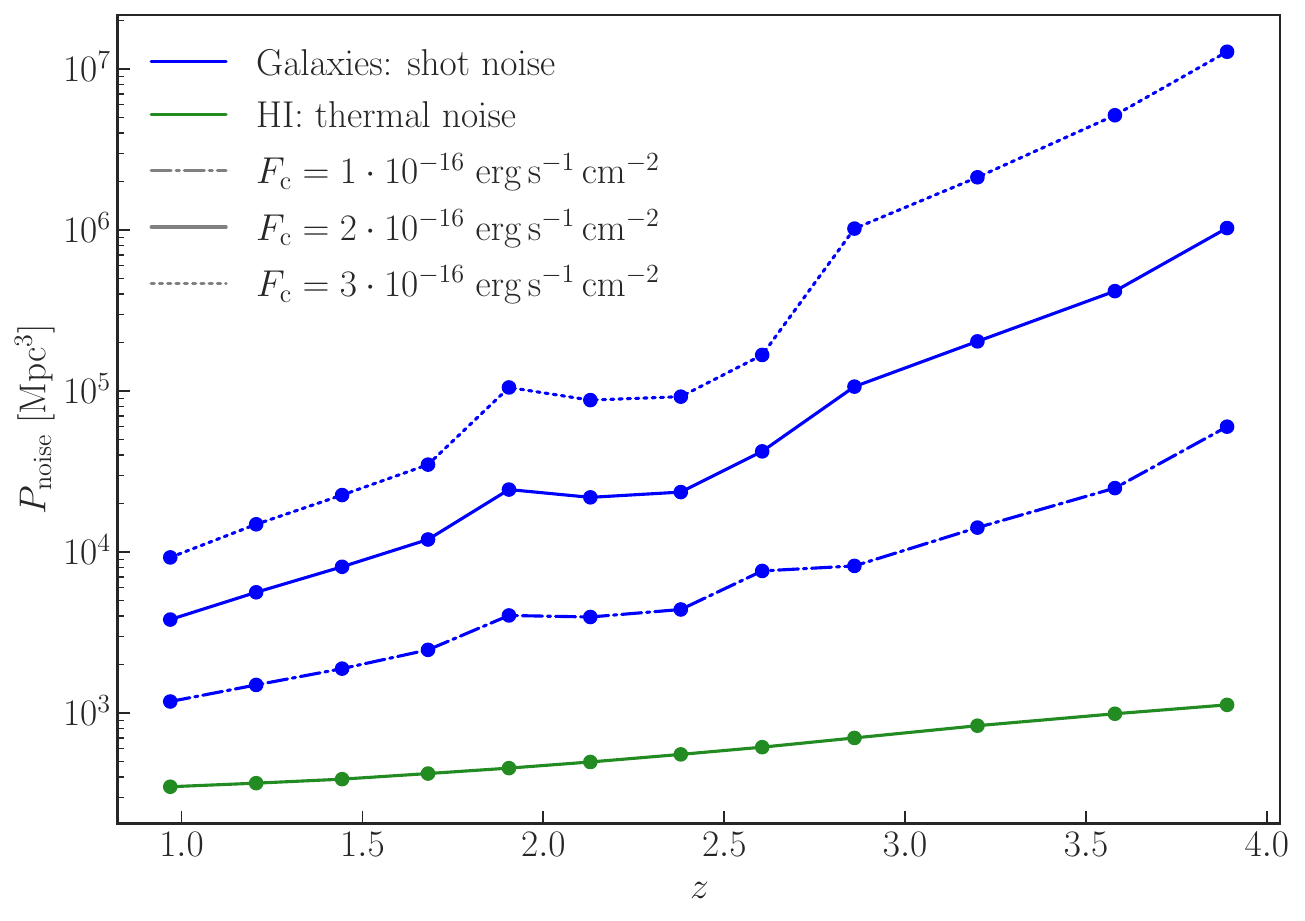}
\caption{\label{noisecomparison}
A comparison between the noise power spectrum for the galaxy survey and the \hi\ intensity mapping survey. Shot noise for the galaxy power spectrum is represented in blue and  is computed for flux limits $F_{\rm c} = 10^{-16} \; {\rm erg \, s^{-1} \, cm^{-2}}$ (dash-dotted), $F_{\rm c} = 2 \times 10^{-16} \; {\rm erg \, s^{-1} \, cm^{-2}}$ (solid) and $F_{\rm c} = 3 \times 10^{-16} \; {\rm erg \, s^{-1} \, cm^{-2}}$ (dotted). Thermal noise, in green, is computed for an \hi\ survey with the specifics given in \cref{HIdata}. The shot noise is larger than the thermal noise at all redshifts, even when lowering the flux limit to $F_{\rm c} = 10^{-16} \; {\rm erg \, s^{-1} \, cm^{-2}}$. }
\end{figure}

\subsection{Cross-correlation between data sets}
\label{ssec:crossdata}
The cross-correlation of two independent tracers of the matter distribution leads to the cross-power spectrum, in which the bias with respect to the matter power spectrum is a combination of the linear biases of both the tracers -- in our case, galaxies and \hi. Concerning the largest scale achievable in this case, we choose to be conservative and to consider as reference the one of the experiment, which can only access the smallest of the two volumes.

Since the observations of spectroscopic galaxies and \hi\ are independent, the galaxy shot noise and the thermal noise in the \hi\ power spectrum are uncorrelated; the variance of the cross power spectrum is therefore just given by the cosmic variance term, while the noise term vanishes, viz.\
\begin{equation} \label{crossnoise}
    P^{\mathrm{noise}}_{\rm g\,\rm\hi}=0 \; .
\end{equation}
In fact, there is in principle a noise term for the cross-correlation, due to the overlap mass range between the dark matter haloes hosting \hi\ galaxies and H$\alpha$ galaxies. However, it has been shown that this term is different from zero only at low redshifts (up to redshift $z \sim 0.5$), and in general it is very small \citep{2015ApJ...812L..22F}. Thus, \cref{crossnoise} is a reasonable assumption for our analysis. As a further test of the robustness of our analysis in this respect, we re-run it using the Fisher matrix formalism and considering a smaller value of $k_{\rm max}$, which would mimic the effect of having small-scale power washed out by the additional cross-correlation noise term. We find that, by fixing $k_{\rm max}=0.05 \; {\rm Mpc^{-1}}$ for all the redshift bins, there is only a 2\% worsening of the marginalised uncertainty on $\fnl$ with respect to the complete case.

Finally, the damping term at large $k_{\perp}$ due to the beam, $\mathcal{D}_{\rm b}$, enters linearly in the cross-power spectrum and, in case \hi\ foreground are treated with the large scale correction given by \cref{fgdamping}, this needs to be taken into account also in the cross-correlation via
\begin{equation}
    P_{\rm g\,\hi}(k,z,\mu) \rightarrow \mathcal{D}_{\rm b}(k,z,\mu) \, \mathcal{D}_{\rm fg}(k,\mu) \, P_{\rm g\,\hi}(k,z,\mu) \; .
\end{equation}

\section{Detection significance}
\label{sec:snr}
First of all, we consider what the overall detection significance of the signal will be. We estimate it via the signal-to-noise ratio (SNR), which  in the multi-tracer scenario reads
\begin{equation}
    {\rm SNR}^{\rm MT}_{\rm overlap}  = \bigg({\,\sum_{k,\mu,z}  {\bm P}^{\sf T}\,\mathsf{Cov}^{-1}({\bm P},{\bm P})\,{\bm P} }\bigg)^{\!1/2}\;,
\end{equation}
where the sum runs over the corresponding $k$-, $\mu$-, and $z$-bins, and `overlap' means that the multi-tracer data vector (including the cross-correlation) is computed considering only the overlapping sky area. It corresponds to the one observed by the experiment with the smallest sky coverage, which in our case is the emission-line galaxy survey, with $f_{\rm sky}\simeq0.36$. Instead, for the auto-correlations of galaxies and \hi\ and their cross-correlation alone, the expression above reduces to
\begin{equation}
    {\rm SNR}^{AB} = \left({\sum_{k,\mu,z}\frac{\left[ P_{AB}\right]^2}{\left[\Delta P_{AB}\right]^2}}\right)^{1/2} \; .
\end{equation}

Since we are considering two surveys with non-perfectly matching sky area, in order not to throw away any data, we include in the full multi-tracer SNR also the contribution of the auto-correlation of the single tracers from the non-overlapping regions \cite{2020JCAP...09..054V}, namely
\begin{equation} \label{snrmttot}
    {\rm SNR}^{\rm MT}_{\rm tot}={\rm SNR}^{\rm MT}_{\rm overlap}  + {\rm SNR}^{ AA}_\textrm{no-overlap} + {\rm SNR}^{BB}_\text{no-overlap} \;. 
\end{equation}

\begin{figure}[tbp]
\centering 
\includegraphics[width=\textwidth]{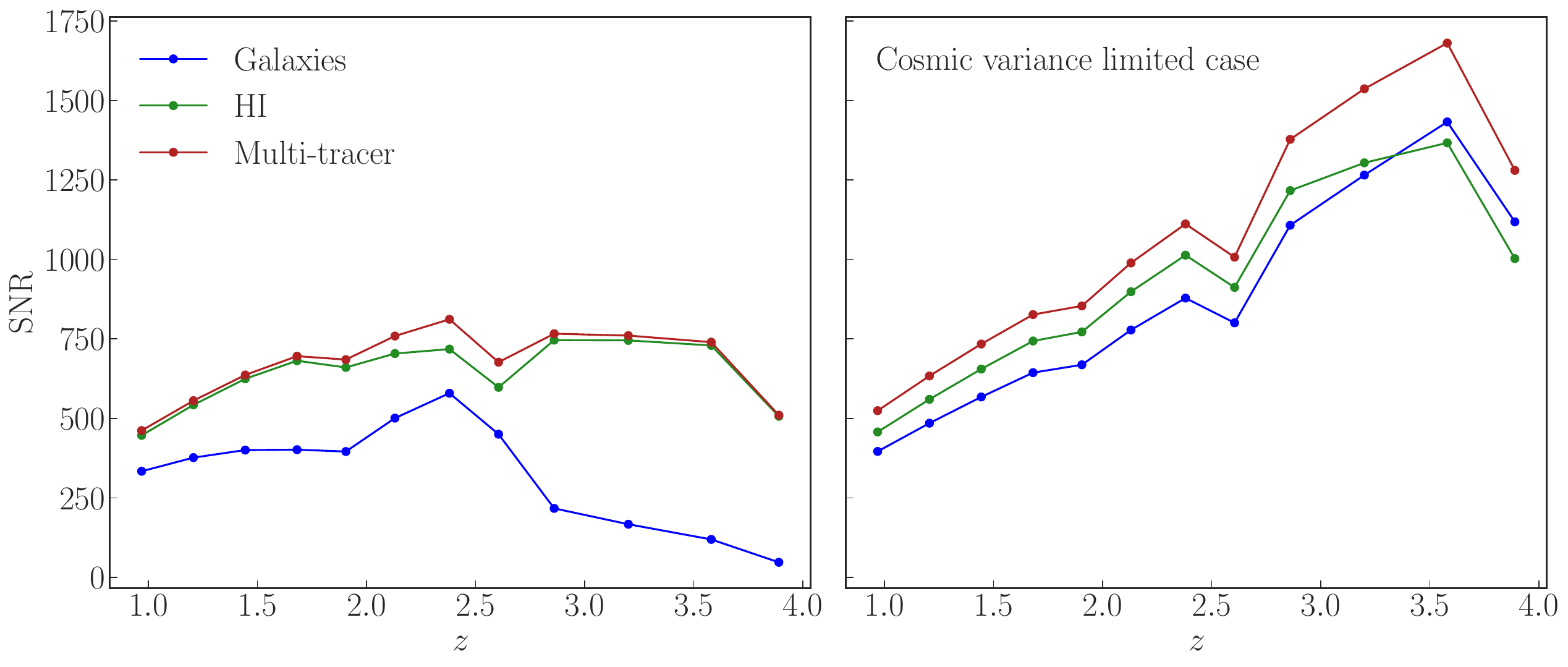}
\caption{\label{snr}
Per-bin SNR  for galaxies (blue),  \hi\ (green), and their multi-tracer (bordeaux). Left panel: All noise terms are included. The SNR of galaxies increases at intermediate redshifts, corresponding to H$\beta$+\oiii\ galaxies, which are an advantageous combination of high linear bias and a still high observed galaxy number density. The SNR of the \hi\ intensity mapping survey is more stable. For the multi-tracer, SNR is driven by the \hi\ term. Right panel: The cosmic variance limited case (zero noise). Breaks in the monotonic trend are related to the spikes in $k_{\rm min}$ due to the transition between different tracers (i.e.\ the impossibility to reach larger scales).}
\end{figure}
This operation lets us retain all the information from the survey with the largest sky coverage, that would be otherwise lost, and it is very advantageous for observation at high redshift and large scales. 

Recalling the expression of the covariance matrix associated to the power spectrum, \cref{eq:covariance}, we can appreciate how the SNR depends on the combination of the effects of the noise term and the number of independent modes, which are different in the case of a galaxy redshift survey or an intensity mapping observation. The shot noise of the galaxy sample and the thermal noise of the \hi\ survey are both scale independent and they increase with redshift. As it can be seen in \cref{noisecomparison}, the shot noise increases much more rapidly than the thermal noise.

Then, the number of independent modes depends on both the probed scale and the redshift; it increases with redshift, whereas it drops the larger the scale. It is also linearly related to the observed sky fraction: when the impact of foregrounds on the \hi\ power spectrum are modelled with the damping factor $\mathcal{D}_{\rm fg}$, we have $f_{\rm sky,g}< f_{\rm sky,\hi}$; viceversa, if the scale cut approach is used, we have $f_{\rm sky,g}> f_{\rm sky,\hi}$ due to the rescaling of the effective volume and observed area. For the \hi, the beam damping and possibly the foreground damping play a role as well, since they result in a loss in the power of the signal. Tracer bias is another relevant element, since the amplitude of the primordial non-Gaussianity correction depends on the term $b_A-1$. This means that if the bias is close to unity, we expect the constraining power on $\fnl$ to be weaker. But we emphasise that this feature is model dependent, as it comes from the ansatz of \cref{unirel}.

\begin{figure}[tbp]
\centering 
\includegraphics[width=\textwidth]{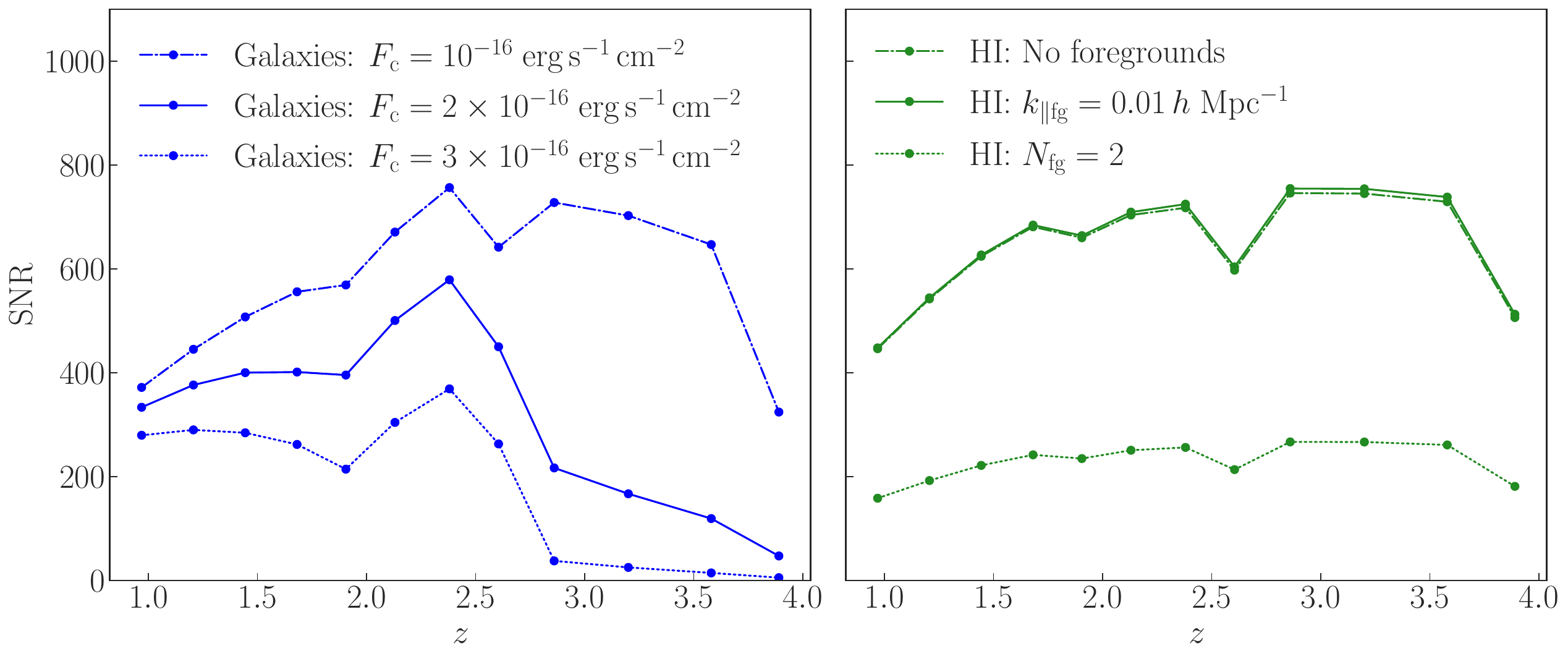}
\caption{\label{snr_tracer}
Left panel:  SNR for the galaxy auto-correlation at different flux limits, $F_{\rm c} = 2 \times 10^{-16} \; {\rm erg \, s^{-1} \, cm^{-2}}$ (solid),  $F_{\rm c} = 3 \times 10^{-16} \; {\rm erg \, s^{-1} \, cm^{-2}}$ (dotted), $F_{\rm c} = 10^{-16} \; {\rm erg \, s^{-1} \, cm^{-2}}$ (dash-dotted). In the last case, since the observed galaxy number density is higher at all redshifts and for all  galaxy types, the increase of SNR for the H$\beta$+\oiii\ sample compared to the other samples is milder. Right panel:  SNR for the \hi\ auto-correlation, computed for different foreground avoidance methods -- the damping factor in \cref{fgdamping} (solid), which is not too dissimilar to the ideal case without foregrounds (dash-dotted line); a scale cut with $N_{\rm fg}=2$ (dotted line) leading to significantly lower SNR.}
\end{figure}

For the benchmark configuration, all these factors together lead to a total SNR that is larger for the \hi\ signal than for galaxies, mainly because of the different sky coverage and the shot noise having more impact than the thermal noise, as depicted in the left panel of \cref{snr}. The cosmic variance limited case is shown in the right panel of \cref{snr}, since it helps to understand the impact of the value of the bias on the SNR and the effects of including the noise terms. As expected, the SNR increases almost monotonically with the redshift, thanks to the inclusion of larger and larger scales and the increasing value of the biases, especially for the different galaxy samples at higher redshift. When comparing this result with the cases where noise terms are included, we notice that there is a general reduction of the detection significance, especially at high redshift, where the noise is more intense and galaxies are more affected (the shot noise is much higher than the thermal noise).

Furthermore, it is interesting to analyse  the responses of the galaxy samples, whose biases range across a wider interval because of the transition between galaxy types. H$\beta$+\oiii\ galaxies are the most robust sample, thanks to the combination of a high bias and an intermediate noise. By contrast, the H$\alpha$ sample, despite its higher number density, has a bias  closer to unity, and the SNR of the \oii\ sample is strongly suppressed by the shot noise. For these reasons, although the multi-tracer SNR in the cosmic variance limited case receives contributions from galaxies and \hi, in the complete case most of the information comes from the \hi\ intensity mapping.

Focusing on galaxies, we also compared the galaxy SNR obtained with different flux limits. The point of interest is the possible balance between variations in the shot noise and in the galaxy bias: for larger values of $F_{\rm c}$, the observed galaxy number density decreases but the galaxy bias is higher, and vice versa. The shot noise is the major source of variations. When considering $F_{\rm c} = 10^{-16} \; {\rm erg \, s^{-1} \, cm^{-2}}$, the SNR for the galaxy survey is almost doubled and becomes comparable with that of the \hi\ intensity mapping survey. By contrast,  for $F_{\rm c} = 3 \times 10^{-16} \; {\rm erg \, s^{-1} \, cm^{-2}}$ the galaxy SNR drops, with  ${\rm SNR}<50$ for $z>2.5$. This behaviour corroborates our previous considerations regarding the impact of the noise term, in particular at $z\geq2.5$. The H$\alpha$ sample partly deviates from this trend and we see a partial mitigation of the effects of the noise. In this case the galaxy number density responds the least to  a change in flux limit, and the SNR is much more sensitive to variation in the galaxy bias, since it is close to unity. Overall, the possible gain due to a more moderate shot noise in the case with $F_{\rm c} = 10^{-16} \; {\rm erg \, s^{-1} \, cm^{-2}}$ is reduced by having a lower bias, while the worsening expected for a less sensitive survey is alleviated by a higher bias.

On the other hand, concerning \hi\ intensity mapping, foregrounds are crucial, as can be seen when moving to the more aggressive foreground removal approach and cutting out the largest scales, or to the ideal situation without foregrounds. In the former case, there is a worsening by a factor of $\sim 3$, while in the latter case the SNR remains almost unchanged. These results, summarised in \cref{snr_tracer}, are shown in the full multi-tracer SNR, which is larger than the  single-tracer SNRs in all cases. 

\section{Analysis}
\label{sec:analysis}
After having assessed the detectability of the signal, the aim of the subsequent analysis is to test to which extent the method is able to detect primordial non-Gaussianity imprints in the tracer power spectrum, estimating the uncertainty on the estimation of $\fnl$. In particular, we focus on comparing the constraining power of the multi-tracer technique with respect to the auto-correlation of a single tracer. We consider different survey configurations, to test their performance and to have a deeper understanding on the impact of different effects, such as the foreground contamination of the 21-cm signal, or the flux limit of a galaxy survey. We analyse the following cases:
\begin{itemize}
    \item Benchmark scenario: Galaxy survey with $F_{\rm c} = 2 \times 10^{-16} \; {\rm erg \, s^{-1} \, cm^{-2}}$ and impact of foreground modelled with the damping factor $\mathcal{D}_{\rm fg}$.
    \item Optimistic scenarios:
    \begin{itemize}
        \item Galaxy survey with $F_{\rm c} = 2 \times 10^{-16} \; {\rm erg \, s^{-1} \, cm^{-2}}$ and no foreground contamination.
        \item Galaxy survey with $F_{\rm c} = 1 \times 10^{-16} \; {\rm erg \, s^{-1} \, cm^{-2}}$ and impact of foreground modelled with the damping factor $\mathcal{D}_{\rm fg}$.
    \end{itemize}
    \item Pessimistic scenarios:
    \begin{itemize}
        \item Galaxy survey with $F_{\rm c} = 2 \times 10^{-16} \; {\rm erg \, s^{-1} \, cm^{-2}}$ and  and impact of foreground modelled with the scale cut using $N_{\rm fg}=2$.
        \item Galaxy survey with $F_{\rm c} = 3 \times 10^{-16} \; {\rm erg \, s^{-1} \, cm^{-2}}$ and impact of foreground modelled with the damping factor $\mathcal{D}_{\rm fg}$.
    \end{itemize}
\end{itemize}

The analysis is performed by sampling the likelihood with Markov chain Monte Carlo algorithms using the publicly available package \texttt{emcee} \cite{2013PASP..125..306F}\footnote{\url{https://emcee.readthedocs.io/en/stable/}}. In analogy with the expression of the signal-to noise-ratio \cref{snrmttot}, we build a Gaussian likelihood function including the multi-tracer combination together with non-overlap single-tracer information, namely
\begin{equation} \label{likelihood}
    \ln \mathcal{L}(\bm d|\bm\theta)^{\rm MT}_{\rm tot} = \ln \mathcal{L}(\bm d|\bm\theta)^{\rm MT}_{\rm overlap} + \ln \mathcal{L}(\bm d|\bm\theta)^{ AA}_\textrm{no-overlap} + \ln \mathcal{L}(\bm d|\bm\theta)^{BB}_\text{no-overlap} \; ,
\end{equation}
where $\bm d$ is the theoretical data vector and $\bm \theta$ an array containing the free parameters of the model, whose best estimate can be obtained by maximising the likelihood.

We consider as free parameters, beside $\fnl$, the primordial spectral index $n_{\rm s}$ and the biases of the tracers, meaning that the terms in \cref{likelihood} providing the single-tracer information do not directly contribute to the reconstruction of the bias of the other tracer; they indeed allow to get globally tighter constraints on the parameters because they do bring relevant information on the other parameters.

We adopt uniform priors on all parameters and assume as fiducial values the \textit{Planck} 2018 cosmology \cite{2020A&A...641A...6P}, and the expected values for the biases \cite{2020MNRAS.493..747P,2018ApJ...866..135V}. We also make sure that the priors are large enough not to bias our results on the recovered values and the posterior distribution. More quantitatively, our priors are several tens of times broader than the final constraints; only in one case being narrower, but still more than three times the size of the final marginalised errors. We decide to keep the bias and the spectral index as free parameters because they are clearly degenerate, albeit to a different extent, with $\fnl$, while we expect the other cosmological parameters to be mostly constrained on smaller scales, or by independent measurements. Note that the amplitude of the primordial fluctuations, \(A_{\rm s}\) is also degenerate with \(\fnl\), and its inclusion in the parameter set may lead to a loosening of the constraints on PNG, if we assume the universality of the halo mass function. However, since cosmic microwave background measurements put tight constraints on \(A_{\rm s}\) from the power spectrum, whereas their constraining power on \(\fnl\) is complementary, coming from the bispectrum, we can safely fix \(A_{\rm s}\) to the values measured e.g.\ by \textit{Planck}.

Hence, our parameter set is \(\bm\theta=\{b_{{\rm g},i},\,b_{{\rm\hi},i},\,n_{\rm s},\,\fnl\}\), with the further subscript `\textit{i}' denoting that the nuisance parameters for the bias(es) correspond to the value of the relevant quantity \textit{in each} redshift bin. Since the biases of the tracers are redshift dependent, for each $z$-bin they correspond to one free parameter in the case of the auto-correlation power spectrum and two free parameters in the case of the cross-correlation power spectrum. This implies a rapidly increasing number of free parameters and a higher computational cost if one wants to exploit the information from more bins jointly. For this reason we first consider two redshift bins at a time (in total four free parameters for the auto-power spectrum, and six for the cross-power spectrum). This will be presented in \cref{ssec:red}.

Then, we also consider bins grouped in order to divide emission-line galaxies of the spectroscopic survey: this allows to improve the amount of information. Moreover the division of the data set according to the transition to different emission-line galaxy types, allow us to evaluate the impact of the galaxy and \hi\ biases on the uncertainty on $\fnl$. Having assumed the universality relation for the assembly bias $b_{\rm \phi}$, the constraining power depends on the amplitude of $b_A-1$. In particular, we expect a more pronounced dependence in the case of the auto-correlation of a single tracer, its bias entering quadratically the power spectrum, whereas the product of the biases of the two tracers in the cross-correlation would mitigate the effects of a sharp variation in the bias of one between galaxies and \hi. This analysis will be discussed in \cref{ssec:elg}.

For the benchmark scenario we also calculate the uncertainty on $\fnl$ considering the whole redshift range.

\section{Results}
\label{sec:results}
\begin{figure}[tbp]
\centering 
\includegraphics[width=.75\textwidth]{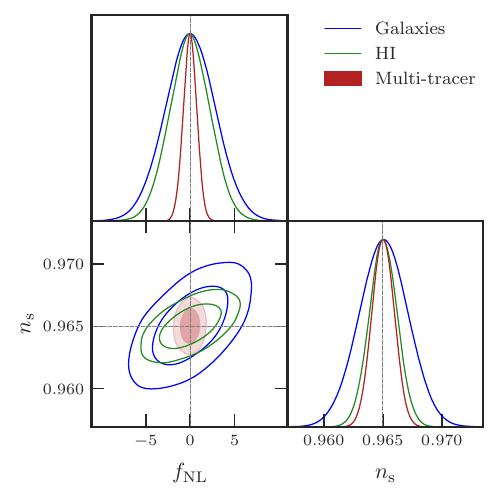}
\caption{\label{contourplot}
Contour plot resulting from the MCMC analysis of the full data set, marginalising over the bias parameters in order to focus on the cosmological ones: blue, green, and bordeaux contours correspond respectively to the posterior distribution of the galaxy auto-correlations, of the \hi\ auto-correlation, and of the multi-tracer technique (this color code will be used in the next plots as well). The multi-tracer technique provides tighter constraints, and, thanks to the combination of independent biased tracers, the degeneracy between $\fnl$ and $n_{\rm s}$ is strongly mitigated.}
\end{figure}

First of all, we present the results of the benchmark configuration analysed spanning over the whole redshift range: the posterior distribution is shown in \cref{contourplot}. The recovered parameters are consistent with the fiducial values whether the analysis is done with the auto-correlation of a single tracer or with the multi-tracer technique. The uncertainty on the primordial non-Gaussianity parameter can be then evaluated by marginalising over all the other parameters. The evaluation of all the redshift range leads to the following constraints on $\fnl$:
\begin{itemize}
    \item Galaxies auto-correlation: $\fnl=0.0 \pm 2.8$;
    \item \hi\ auto-correlation: $\fnl=0.0 \pm 2.3$;
    \item Multi-tracer technique: $\fnl=-0.01 \pm 0.76$.
\end{itemize}

\hi\ constraints are slightly tighter than those provided by galaxies because the thermal noise is lower than the shot noise and also because, in this setup, the \hi\ intensity mapping survey would cover a larger cosmic volume and this results to be advantageous even in presence of foregrounds, if their effect is modelled as a damping factor in the 21-cm signal. 

The multi-tracer technique allows us to get an uncertainty on $\fnl$ that is more than halved with respect to those coming from the simple auto-correlation, despite the bigger amount of free bias parameters (a total of 26 free parameters for the multi-tracer technique, instead of the 14 needed for the auto-correlations alone). Moreover, reaching $\sigma (\fnl) \simeq 1$ is crucial: many inflationary models predict $\fnl \simeq 1$, making $\sigma (\fnl) \simeq 1$ a relevant threshold to get to in order to precisely constrain this parameter \cite{2017PhRvD..95l3507D,2014arXiv1412.4671A}.

\subsection{Redshift dependence}
\label{ssec:red}
\begin{figure}[tbp]
\centering 

\includegraphics[width=\textwidth]{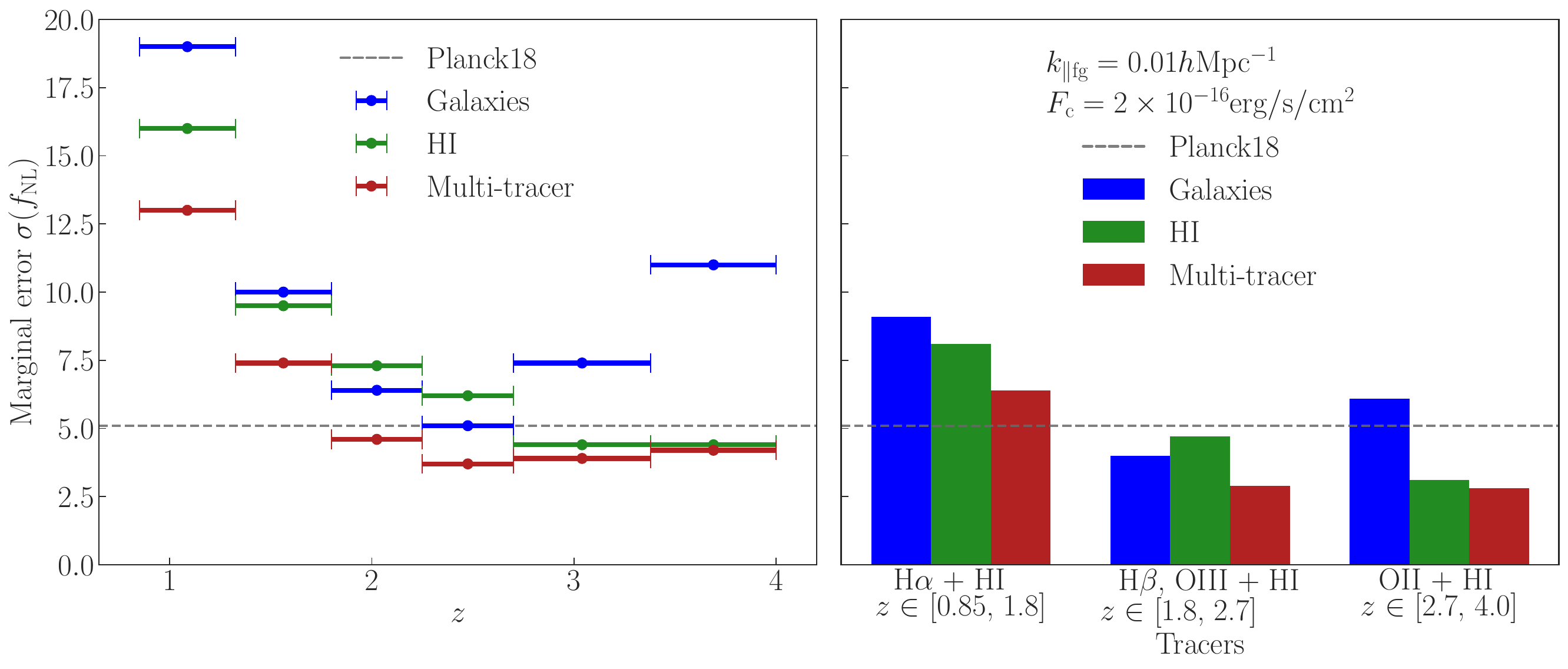}
\caption{\label{constr_fgdampf2}
Left: $68\%$ error bars on $\fnl$ from the analysis of two redshift bins at a time. Horizontal bars represent the total width of two bins considered. Right:  Constraints  from the analysis with redshift bins grouped in order to follow the division in galaxy types. For comparison, the constraint  from the cosmic microwave background \cite{2020A&A...641A...9P} is shown by the dashed grey line. }
\end{figure}

By analysing the redshift bins two by two, it is possible to evaluate how the constraining power of this analysis varies as a function of the redshift. The primordial non-Gaussianity parameter fiducial value is always efficiently recovered with an uncertainty that decreases going to higher redshifts, as a consequence of the fact that larger volumes can be accessed at high $z$. This is shown in the left panel of \cref{constr_fgdampf2}.

Focusing on the auto-correlation of the galaxy sample, it can be noticed that $\fnl$ is not well constrained in the low-$z$ bins, both because very large scales cannot be accessed at those redshifts, and because the bias term $(b_{\rm g}-1)$ is very small for the galaxies at those redshifts. This, in turn, lowers the effects of $\fnl$ of the power spectrum. On the other hand, at high redshift, in the bins corresponding to the \oii\ sample, the uncertainty on $\fnl$ saturates and, after reaching a lower limit, it slightly increases; this is due to the fact that the shot noise becomes too large and it starts to dominate the observed power spectrum. Thus, the advantage of accessing large scales is lost. A test was done considering a configuration in which the galaxy power spectrum is assumed to be cosmic variance limited, thus without shot noise: the constraints obtained on $\fnl$ improve at every redshift but in particular at high redshifts. These effect are not present in the auto-correlation of the \hi\ signal: the higher SNR and the larger sky coverage (despite the foregrounds) ensure a better constraining power for the \hi\ auto-correlation. 

The multi-tracer technique results in be the best performing method to constrain primordial non-Gaussianity. It is not affected  by the low SNR at high $z$ of the power spectrum of galaxies, nor by its limited scale range at low $z$, nor by the loss of information due to foregrounds contaminating the \hi\ signal. The improvement on the constraints is up to 30\% with respect to the best performing tracer in the corresponding bins. The gain is lower when one of the two tracers is individually performing significantly better: in this case that tracer gives the dominant contribution in the multi-tracer technique.

The constraining power on the others free parameters of the multivariate analysis is also taken into account. The primordial spectral index and the bias parameters are all recovered within a confidence level of $68\%$ in both the auto-correlation of the single tracers and the multi-tracer technique, even in the redshift bins where galaxies cannot constrain $\fnl$. When applying the multi-tracer technique, the uncertainty associated to these parameters is reduced as well, mostly as a consequence of reduced degeneracies between parameters.

\subsection{Tracer dependence}
\label{ssec:elg}
The analysis of more redshift bins at a time leads to a global improvement of the results, as displayed in the right panel of \cref{constr_fgdampf2}: the constraints on $\fnl$ are reduced up to a factor $\sim 2$ thanks to the improvement of the amount of information, and there is a global gain in the performance also concerning the other free parameters. As in the analysis performed per redshift bins, $\fnl$ is recovered with an uncertainty decreasing with increasing redshift.

In the present analysis, the differences between the various tracers are highlighted, in particular concerning the bias term. The H$\alpha$ galaxies still suffer from the scale limit imposed by the small surveyable volume at low redshift, as well as from the small bias. As a consequence, the constraints from their auto-correlation are less competitive. The best performing sample is the H$\beta+$\oiii\ sample, thanks to a combination of three advantageous conditions: first of all the effects of a primordial non-Gaussianity are enhanced by a large bias; secondly, it is still possible to observe a significant number of galaxies; and, lastly, the scale range reachable at those redshifts is extended thanks to much larger volume. For the \oii\ sample only two of those conditions are met: the large bias and the progressive enlargement of the scale range, however this sample is affected by its low density, which prevents the constraints to improve with respect to the previous sample.

Since in the case of the \hi\ intensity mapping, there is no transition between different samples, the considerations on the effects of the bias are less relevant because the trend of $b_{\rm\hi}$ is smooth and moreover $b_{\rm\hi}$ significantly deviates from unity even at low $z$. Thus the results are mainly driven by the accessible scale range, which is wider at high $z$, so that $\fnl$ is recovered with decreasing uncertainty at high redshift. 

All such considerations remain valid when the multi-tracer technique is applied, but the results are more solid against possible degrading effects such as the galaxy shot noise or the small bias term. The gain with respect to the simple auto-correlation is comparable to what can be obtained with the analysis with only two redshift bins: to get a larger improvement more information should be added analysing all the redshift range available.

\subsection{Flux dependence}
\label{ssec:flim}
By varying the flux limit of the galaxy survey, we can appreciate the impact of the shot noise on the final result.

The variation of the observed number density with flux cut is more evident for the oxygen-line samples, namely the H$\beta$+\oiii\ sample and the \oii\ sample; and especially for this last case, since it is the one at the highest redshift, the detectability of the sources is therefore very responsive to any change in the specifics of the survey. When varying the flux limit, the galaxy bias needs to be evaluated according to it. 

\begin{figure}[tbp]
\centering 
\includegraphics[width=\textwidth]{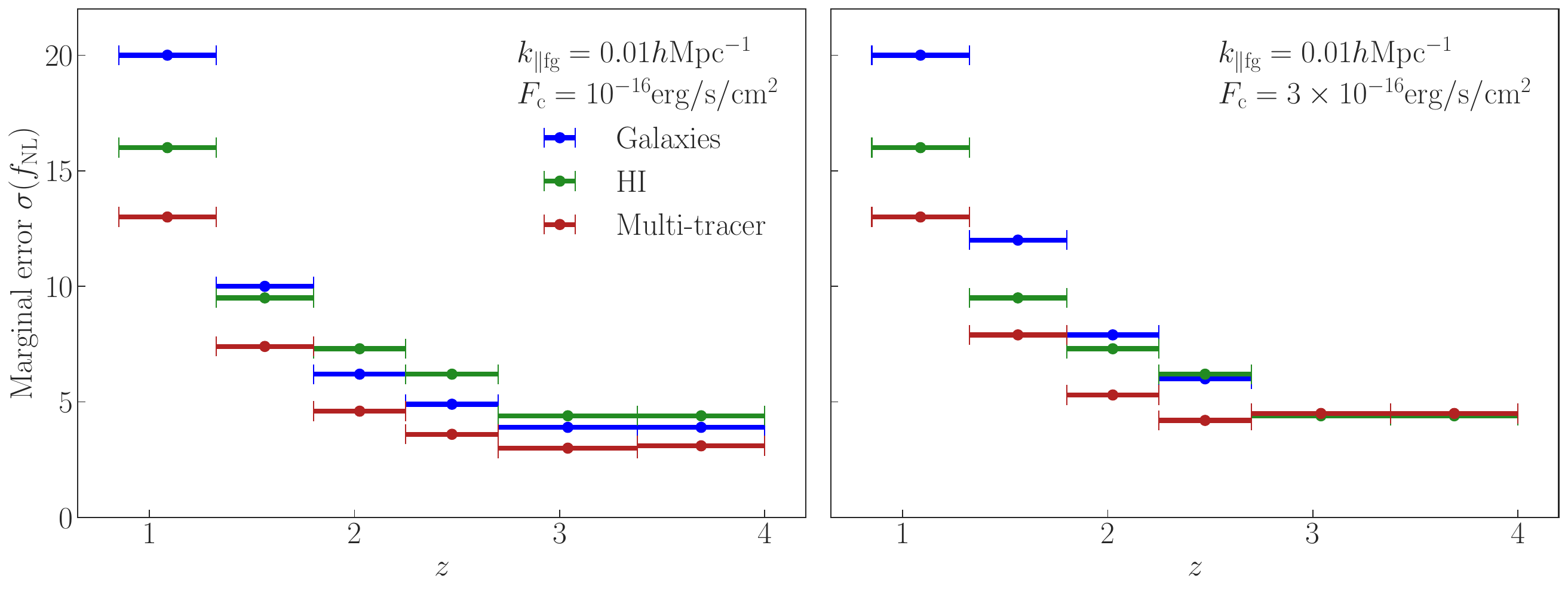}
\vfill
\includegraphics[width=\textwidth]{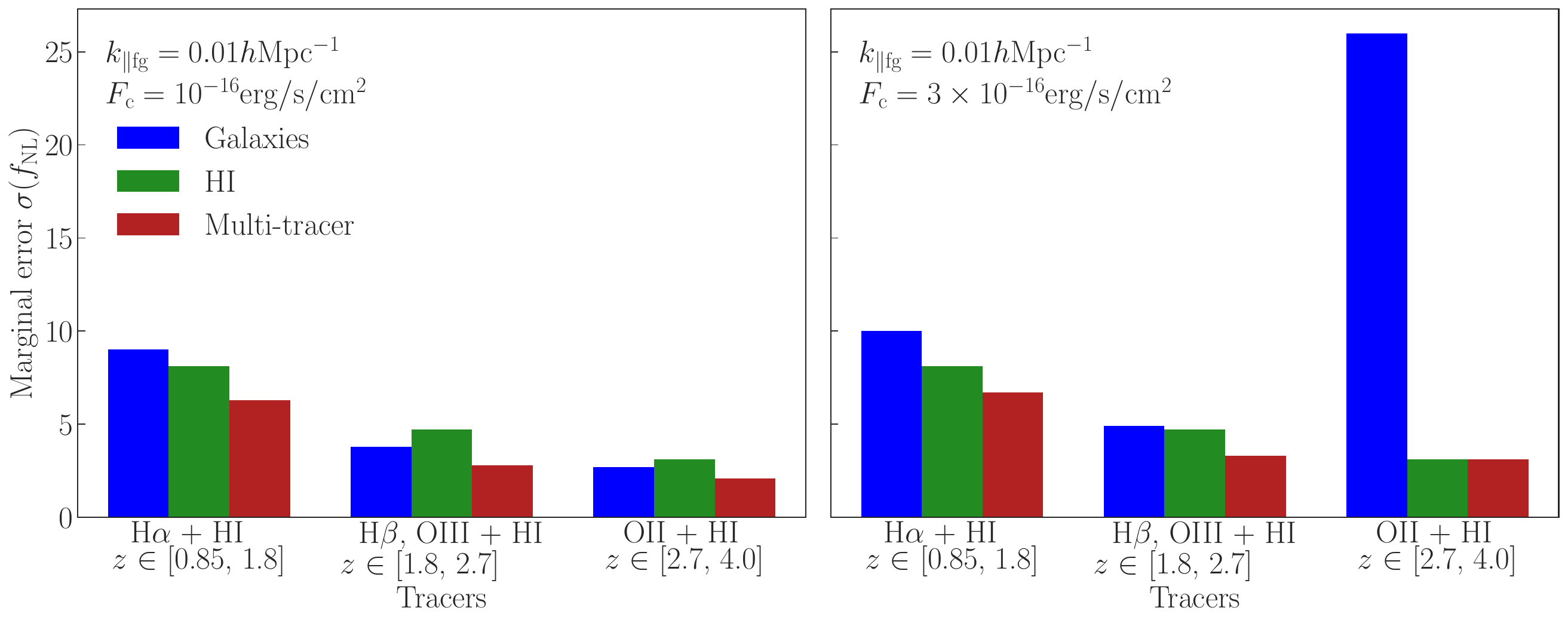}
\caption{\label{constr_flux}
Marginalised uncertainty on $\fnl$ for an optimistic scenario with $F_{\rm c} = 10^{-16} \; {\rm erg \, s^{-1} \, cm^{-2}}$ (left panels) and for a pessimistic scenario with  $F_{\rm c} = 3 \times 10^{-16} \; {\rm erg \, s^{-1} \, cm^{-2}}$ (right panels). Top panels refer to the analysis discussed in \cref{ssec:red} (as in \cref{constr_fgdampf2} left). Bottom panels refer to the analysis discussed in \cref{ssec:elg} (as in \cref{constr_fgdampf2} right). Foregrounds in the \hi\ intensity mapping are treated with the benchmark approach with the damping factor described in \cref{fgdamping}, so that the uncertainty on $\fnl$ from the \hi\ auto-correlation is the same as in \cref{constr_fgdampf2}.}
\end{figure}

The results are shown in \cref{constr_flux}. In general, they do not change significantly at low $z$, where constraining power is limited by having a very small term $(b_{\rm g}-1)$ and a small volume. The main differences arise at high redshift both in the analysis with two or four redshift bins, mostly because of the variation of the shot noise.

Considering the optimistic scenario of a survey with $F_{\rm c} = 1 \times 10^{-16} \; {\rm erg \, s^{-1} \, cm^{-2}}$, all the parameters are well constrained and focusing on $\fnl$ the uncertainty decreases with respect to the benchmark configuration presented in \cref{ssec:red,ssec:elg}. The galaxy auto-correlation leads to constraints which are up to a factor of three tighter and the multi-tracer technique as well takes advantage from this improvement. Notably, the uncertainty on $\fnl$ does not saturate anymore at the redshifts corresponding to the \oii\ sample, which is therefore able to provide stronger constraints.

Moving to the more stringent scenario with $F_{\rm c} = 3 \times 10^{-16} \; {\rm erg \, s^{-1} \, cm^{-2}}$, the performance of the analysis resents the higher shot noise, which for the \oii\ sample is always above the cosmological signal. This implies that the posterior distribution for $\fnl$ is prior dominated for $z\gtrsim2.7$ in the case of the galaxy auto-correlation, and it is driven by the \hi\ signal in the case of the multi-tracer approach. The degradation of the constraints on $\fnl$ is less severe at intermediate $z$, corresponding to the H$\beta+$\oiii\ galaxies: despite the higher value of $F_{\rm c}$, this sample is still dense enough and the higher bias consents to precisely recover the primordial non-Gaussianity parameter. 

For the SNR, the combination of the variation of the bias and of the shot noise deserves deeper investigation. It can be noticed that the balancing between the two competing terms becomes more effective for the H$\alpha$ sample. In particular,  there is full compensation in the first redshift bins. Finally, it is worth highlighting that, even if the flux limit has a great impact on the performance of the galaxy auto-correlation, the multi-tracer approach is less sensitive to the variation of the flux limit of the survey.

\subsection{Foregrounds}
\label{ssec:fg}
Foregrounds are one of the main challenges in the detection of the \hi\ cosmological signal and the evaluation of their impact is a key point when constraining cosmological parameters. The consequences of the loss of signal due to foregrounds can be estimated considering the ideal case in which the \hi\ signal is foreground-free. In this case, with the analysis of \hi\ in auto-correlation the cosmological parameters and the bias parameters are recovered to their fiducial values with higher accuracy because there is no damping in the signal at large scales; the mean improvement is around $15\%$ with respect to our standard scenario, with the biggest gain at low redshift. As a consequence of the larger amount of information especially at large scales, the multi-tracer technique provides tighter constraints on the primordial non-Gaussianity parameter as well.

\begin{figure}[tbp]
\centering 
\includegraphics[width=\textwidth]{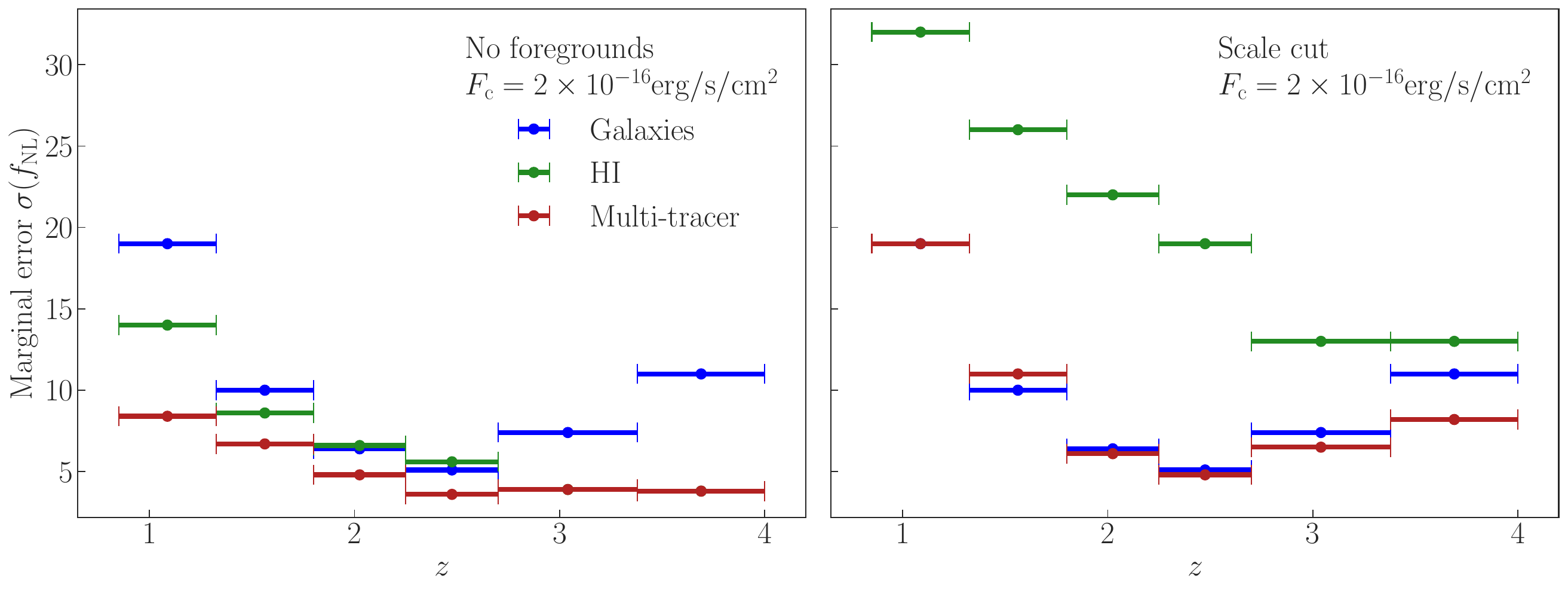}
\vfill
\includegraphics[width=\textwidth]{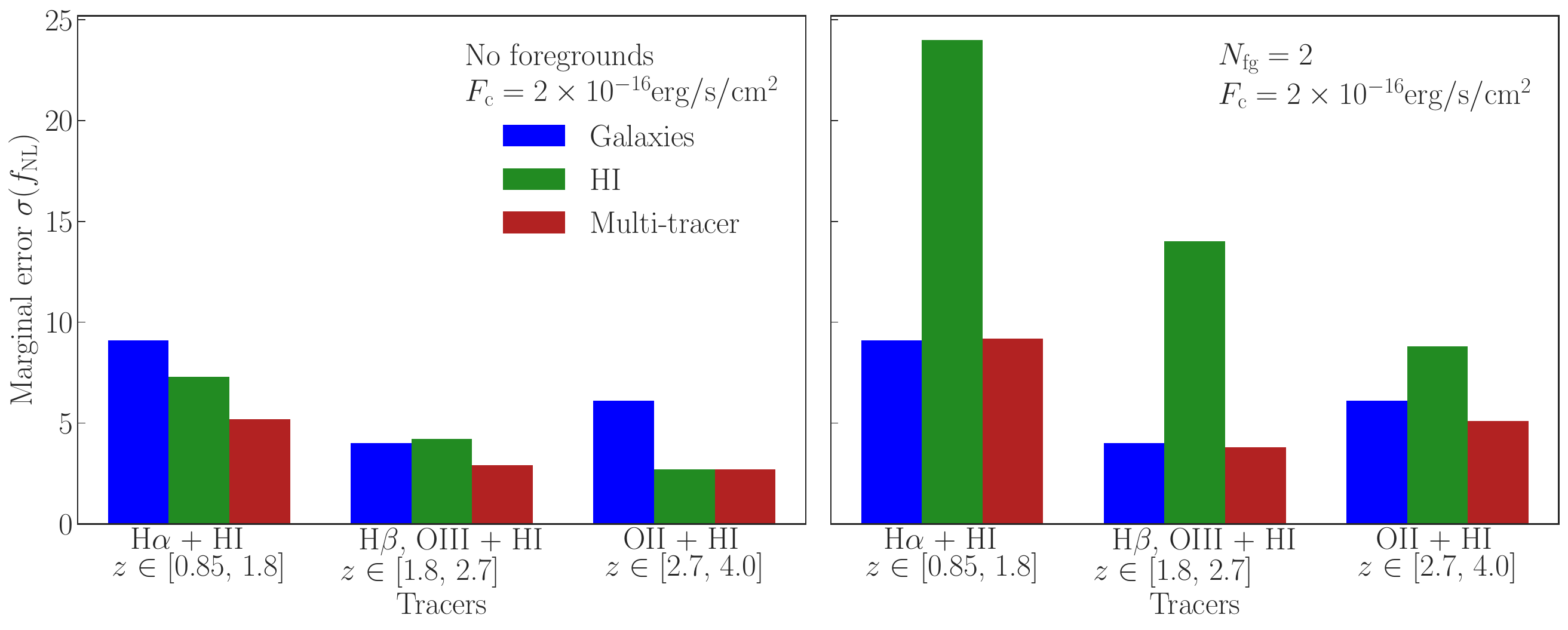}
\caption{\label{fig:constr_fg}
Impact of foreground avoidance on the uncertainty on $\fnl$. Left panels show forecasts without foregrounds, while right panels include a scale cut in the \hi\ intensity mapping. Top panels refer to the analysis discussed in \cref{ssec:red} (as in \cref{constr_fgdampf2} left), while bottom panels refer to the analysis discussed in \cref{ssec:elg} (as in \cref{constr_fgdampf2} right). The flux limit of the galaxy survey is kept constant at $F_{\rm c} = 3 \times 10^{-16} \; {\rm erg \, s^{-1} \, cm^{-2}}$, so the results for galaxy auto-correlation are the same as in \cref{constr_fgdampf2}.}
\end{figure}

On the other side, when adopting a more strict approach to avoid foregrounds, the loss of signal can be critical. If one needs to exclude all the modes below a given threshold, as explained in the last part of \cref{HIdata}, the analysis suffers from a lack of information on those scales which are more sensitive to the presence of primordial non-Gaussianity. The worsening of the results is significant at all $z$ and in the case of the auto-correlation of \hi\ the uncertainty on $\fnl$ is at least doubled with respect to the reference configuration. Thanks to the combination of galaxies and \hi\ in the multi-tracer technique the degradation of the constraints is mitigated.

\section{Conclusion}
\label{sec:conclusion}
In the next years cosmological surveys aiming to map the large-scale structure of the Universe will shed light on the characteristics of the primordial Universe, and thanks to the development of the \hi\ intensity mapping technique, the complementarity of two observables -- galaxies in the optical band and \hi\ in the radio band -- will be a powerful tool to exploit.

In this context, we presented a forecast on the performance in measurements of the primordial non-Gaussianity parameter, $\fnl$, done in a Bayesian framework, to assess the gain on the constraining power that can be reached when applying the multi-tracer technique. The scale dependent bias induced by the presence of a primordial non-Gaussianity leads to a change in the slope of the power spectrum of a matter tracer, mostly evident at large scales, where a multi-tracer approach is expected to grant the best performance thanks to the possibility of beating the cosmic variance.   

Starting from a fiducial cosmology generated with \texttt{CAMB}, we simulated a synthetic data set for an emission-line galaxy survey, for which we also took into account interloper samples at high $z$ in order to extend the redshift coverage. Concerning the \hi\ intensity maps, we simulated the signal including instrumental effects, such as the beam damping, and the loss of signal at large scales due to astrophysical foregrounds.

By means of MCMC methods, we evaluated the constraining power of the analysis in terms of the accuracy on the reconstruction of our free parameters ($\fnl$, $n_{\rm s}$, and the biases of the tracers), focusing on the primordial non-Gaussianity parameter. The full multi-tracer likelihood was build including not only the data from the overlapping region between the galaxy survey and the \hi\ intensity mapping survey, but also what the individual tracers provide from non-overlapping regions, allowing us to retain all the available information, especially at large scales.

We found that this method is unbiased and that all the parameters are recovered to their fiducial value both with the single-tracer and the multi-tracer approach. However, when performing the analysis on the total data set, the multi-tracer technique provides an improvement of a factor greater than 2 on the uncertainty associated to the reconstructed parameters and especially on $\fnl$, showing that it will be advantageous to go for a multi-tracer approach and therefore design surveys in different bands with maximised overlapping coverage both in redshift and in the observed sky patch. 

When considering the results with respect to the redshift, it is evident that the constraining power improves when going to high redshift, when larger cosmic volumes are accessible, as long as the noise term does not exceed the cosmological signal. This is the case of the last redshift bins of the galaxy sample, where the observed number density of \oii\ interlopers is insufficient to provide tight constraints on $\fnl$. This does not mean however that this sample must be excluded, because the multi-tracer technique results to be quite robust against the increase of the noise of one of the samples considered. Beside the effect of the scale range, we assessed the impact of the amplitude of the bias, which can be mostly appreciated from the analysis of the galaxy auto-correlation: the H$\alpha$ sample, despite being the main sample of the modelled galaxy survey, preforms the worst even compared with the interloper samples precisely because of the values of the bias. This justifies again the need of exploiting also other samples beside the target one. 

A caveat must be added in the discussion of the bias, since we used the universality relation for $b_{\rm \phi}$. Recent literature demonstrates that this might not be the correct approach, and that further investigation of the assembly bias is needed \cite{2023JCAP...01..023L,2020JCAP...12..031B,2023arXiv230209066B,2020JCAP...12..013B}. Identifying a better model for $b_{\rm \phi}$  would allow us to achieve more accurate results on primordial non-Gaussianity. However, the multi-tracer technique provides some mitigation here, since it is more robust  to inaccurate models of the assembly bias \cite{2023arXiv230209066B}. 

We evaluated the impact of the shot noise associated to the galaxy power spectrum by varying the flux limit of the survey, finding that the main differences arise at high redshift and in the analysis of the galaxies in auto-correlation, while the multi-tracer technique is more stable against variation in the observed galaxy number density. Concerning the \hi\ intensity mapping data, foregrounds were taken into account with different approaches in order to estimate how much the constraints on $\fnl$ are affected by how the foreground avoidance is modelled. We found that there is a high variability in the results depending on how severe we model the loss of signal to be. The development of techniques to clean the \hi\ intensity mapping data from foregrounds without erasing the cosmological signal is therefore crucial: blind and non-parametric methods are typically employed \cite{2020MNRAS.499..304C,2021MNRAS.504..208C,2023MNRAS.523.2453C,2022MNRAS.509.2048S} and it will be important to explore in detail the best settings, eventually in combination with techniques to reconstruct the \hi\ signal. 

Globally, the constraints we found from our forecast on the presence of a primordial non-Gaussianity are tight, reaching their best value thanks to the multi-tracer technique. These are promising results, suggesting that next-generation survey both in the optical and the radio bands will allow to probe inflationary models and the dynamics of the early Universe with high accuracy. In order to do this, we should take advantage of the multi-tracer technique to overcome the cosmic variance.

\acknowledgments
MBS and SC warmly thank Isabella Carucci for valuable discussions in the early stages of this work, and Federico Montano for support in the computation of the linear bias for different emission-line galaxy samples. SC also acknowledges support from the `Departments of Excellence 2018-2022' Grant (L.\,232/2016) awarded by the Italian Ministry of University and Research (\textsc{mur}) and from the Italian Ministry of Foreign Affairs and International
Cooperation (\textsc{maeci}), Grant No.\ ZA23GR03. RM is supported by the South African Radio Astronomy Observatory and the National Research Foundation (Grant No.\ 75415). 
\appendix
\section{Some robustness tests} \label{appendix:apa}

The results of the forecasts presented depend on some assumptions whose impact deserves  further investigation, in order to assess the robustness of the analysis and the multi-tracer method.

\begin{figure}[tbp]
\centering 
\includegraphics[width=.75\textwidth]{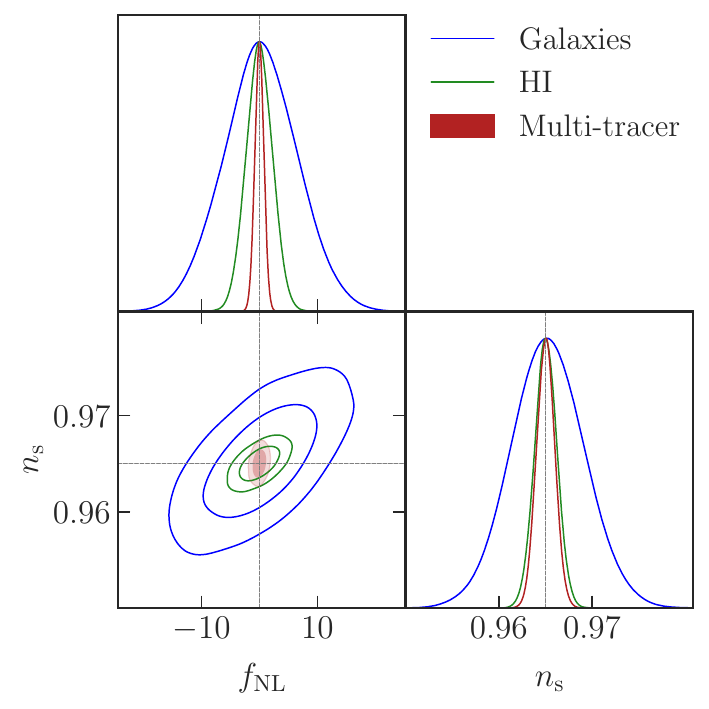}
\caption{\label{contourrom}
Contour plot resulting from the MCMC analysis of the \rom-SKAO data set, marginalising over the bias parameters. If compared with \cref{contourplot}, the widening of the galaxy auto-correlation contours is the most evident difference, whereas there is only a slight difference in the  multi-tracer contours. The multi-tracer  is confirmed to be more constraining than both single tracers in auto-correlation and also to break the degeneracy between $\fnl$ and $n_{\rm s}.$
}
\end{figure}

A variation of the sky coverage of a survey is the first aspect we explored, since it impacts the largest scales observable, where  the effects of  $\fnl$ are strongest. For this reason, we repeated the full MCMC analysis considering a \rom-like galaxy survey (i.e.\ covering a smaller patch of the sky but deeper in flux) \citep{2022ApJ...928....1W} and keeping as a benchmark the \hi\ intensity mapping survey, since the SKAO will dominate the radio panorama. 
Specifically, we considered a galaxy survey observing $2\,000\,\deg^2$ in the sky with a flux limit of $F_{\rm c} = 10^{-16} \; {\rm erg \, s^{-1} \, cm^{-2}}$, while we left unchanged the redshift range, assuming that it is covered thanks to the inclusion of the interloper samples. The results in \cref{contourrom} show that the loss of large-scale information from the galaxy survey, even with a higher galaxy number density, degrades the results of the galaxy single-tracer analysis, where all the parameters are constrained but with a larger uncertainty. By contrast, there is no significant impact on the multi-tracer analysis, thanks to the inclusion of the non-overlapping \hi\ signal in the likelihood. More quantitatively, there is a few percent worsening of the uncertainty on $\fnl$, $n_{\rm s}$, and the galaxy biases (but no difference in the \hi\ biases, since the  signal is unaffected).

The possibility of including interloper samples in order to extend the redshift range covered with a galaxy survey is another key point of these forecasts and it relies in turn on two assumptions. Firstly, the interloper identification is feasible and efficient, so that their observed number density is high enough to prevent the shot noise to dominate the cosmological signal. Secondly, their bias can be properly modeled. To investigate the impact of these aspects, we used a Fisher matrix formalism and analysed them one at a time while keeping all the other variables (flux limit, sky coverage, etc.) as in the benchmark configuration. 

On the number density of interlopers, we recall \cref{galng}, where the flux limit of the survey is taken into account when integrating the luminosity function to find ${\bar n}(z)$, meaning that only the objects that appear bright enough can be detected. This already implies that the observed galaxy number density is expected to be lower at high redshift, in the bins that we assume to cover with interloper line galaxies, since $L_{\rm min}(z)$ scales with the square of the luminosity distance at a given flux limit. In regard to the completeness of the interloper samples, it is not guaranteed that all the interlopers above the flux limit are included in the number count, and this can lead to a reduced observed number density.

\begin{figure}[tbp]
\centering 
\includegraphics[width=.85\textwidth]{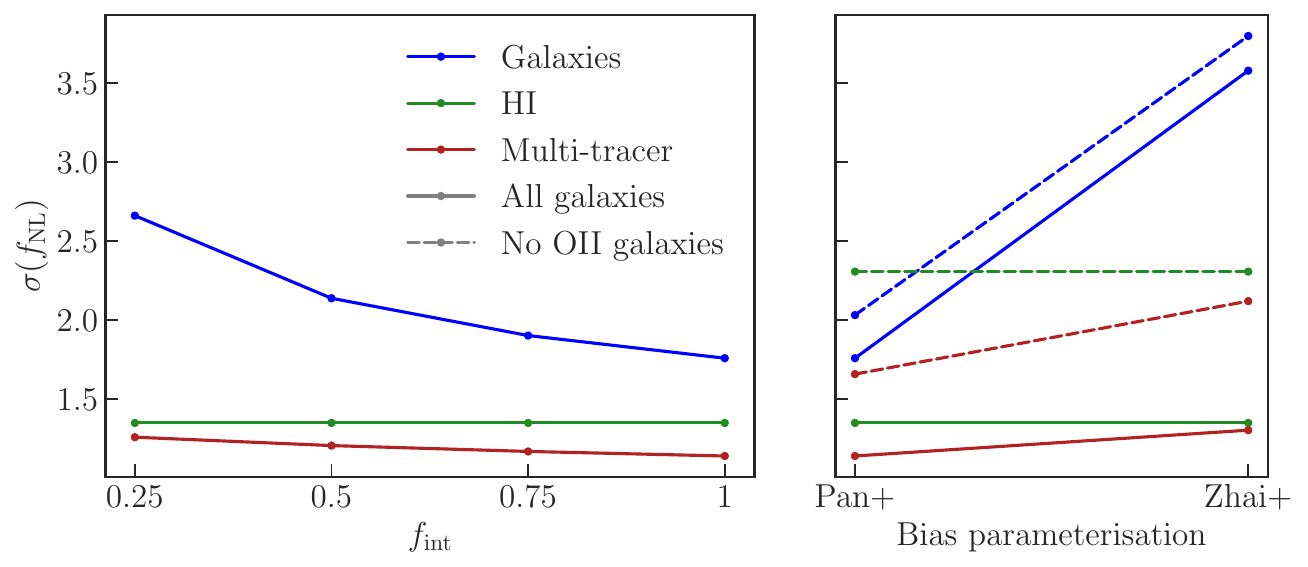}
\caption{\label{niter_bias}
Left panel: 68\% error bars on $\fnl$ for  $f_{\rm int}$ from 0.25 (a quarter of the observable interlopers  detected) to 1 (all detected). The blue line (galaxy auto-correlation) is much steeper than the bordeaux line (multi-tracer):  a depletion of the interloper samples downgrades the  galaxy single-tracer results, but does not  significantly affect the multi-tracer results. The green line (\hi\ auto-correlation) is shown as reference. Right panel: 68\% error bars on $\fnl$ obtained with different bias models -- considering the whole redshift range (solid lines) or excluding the last redshift bins (dashed lines). The galaxy auto-correlation is more sensitive to variation of the value of the bias, whereas the multi-tracer constraints show only a few percent change. Note that removing the redshifts above $z=2.6$ has a different impact on the \hi\ and galaxy single-tracer  and on the multi-tracer analysis, but the overall trends are preserved.
}
\end{figure}

In order to understand how the constraints on $\fnl$ respond to the possible incompleteness of the samples of H$\beta$+\oiii\ and \oii\ galaxies, we perform a Fisher forecast. We re-scaled the number density of interlopers in order to consider only a fraction $f_{\rm int}$ of the full samples, while keeping fixed the H$\alpha$ number density, i.e.,
\begin{equation}
    {\bar n}(z) \to f_{\rm int}\,{\bar n}(z) \; .
\end{equation}
As depicted in the left panel of \cref{niter_bias}, this leads to a gradual broadening of the constraints on $\fnl$ obtained with the galaxy auto-correlation, up to a $\sim50\%$ worsening in the case that only $25\%$ of interlopers is detected. On the other hand, the multi-tracer  uncertainty on $\fnl$  is at most only $10\%$ larger than in the benchmark scenario. This confirms the stability of the multi-tracer technique even against a degradation of the quality of the data sets, thanks to the combination of more than one tracer --  also in agreement with the results in \cref{ssec:flim}.

\begin{figure}[tbp]
\centering 
\includegraphics[width=.65\textwidth]{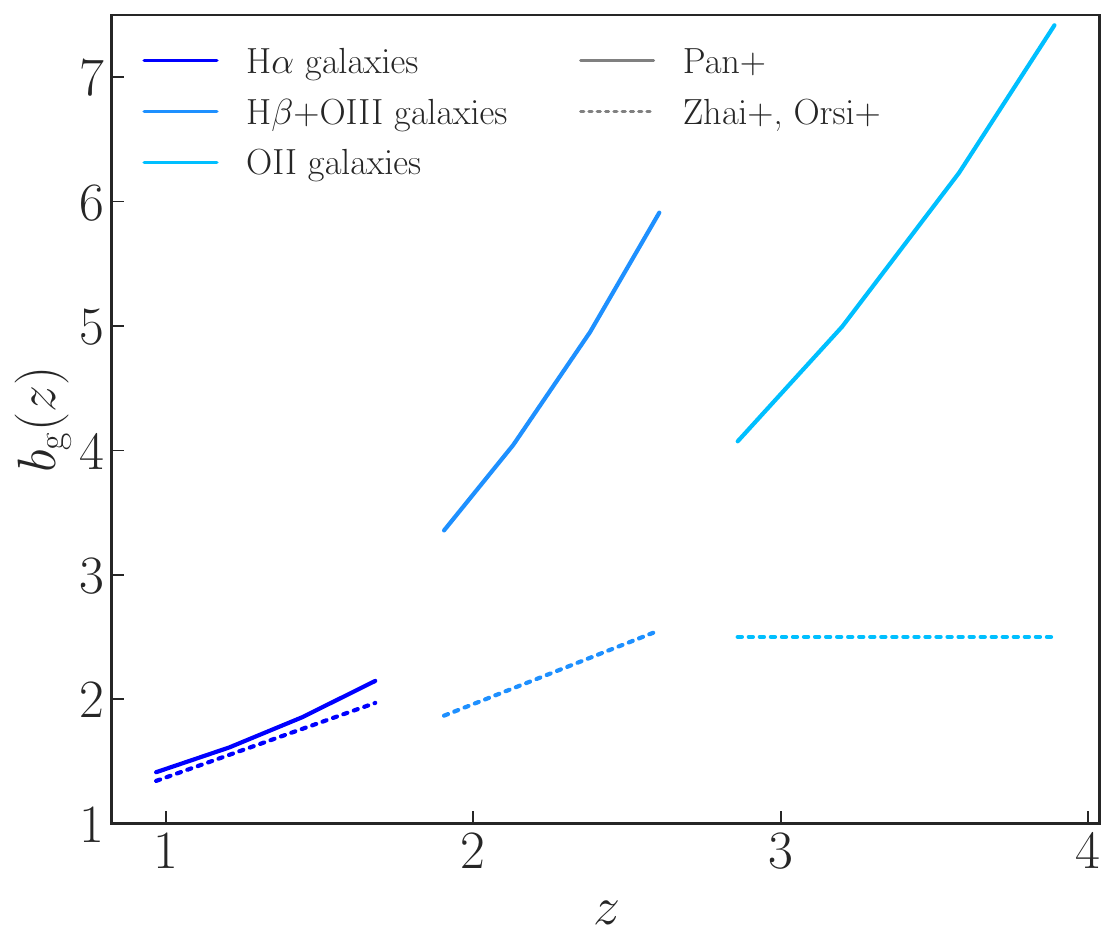}
\caption{\label{bias_mod}
Galaxy bias computed with various parametrisations for the different galaxy types. Solid lines correspond to the fitting formula of \cite{2020MNRAS.493..747P},  for a flux limit  $F_{\rm c} = 2 \times 10^{-16} \; {\rm erg \, s^{-1} \, cm^{-2}}$. Ddashed lines correspond to the models in \cite{2021MNRAS.505.2784Z} (H$\alpha$ and \oiii\ galaxies) and in \cite{2014MNRAS.443..799O} (\oii\ galaxies). 
}
\end{figure}

The  strength of the features due to primordial non-Gaussianity is related to the value of the bias through the $b_A-1$ term entering the expression of the scale-dependent bias. This means that any inaccuracy in the model of the bias could be reflected in the constraints on $\fnl$. Among the different galaxy samples and \hi, the interloper samples are again the ones whose modeling is the most difficult to describe, so we explored a different parametrisation of the bias of interloper galaxies at high redshift. We adopted the parametrisation of \cite{2021MNRAS.505.2784Z} for the intermediate sample (and for consistency also for the H$\alpha$ sample) and, following \cite{2014MNRAS.443..799O}, we used a fixed bias of $b_{\rm g}=2.5$ for the \oii\ sample, as shown in \cref{bias_mod}. The main goal of this analysis is to compare the results from the bias model of \cite{2020MNRAS.493..747P}, which shows a steep evolution in redshift, and from other models, where the galaxy bias assumes smaller values even at high $z$. Hence, even a fixed value for the \oii\ galaxies can at first approximation be used to highlight this difference. We also emphasise that the fitting formula in \cite{2021MNRAS.505.2784Z} is obtained considering a flux limit of $F_{\rm c} = 10^{-16} \; {\rm erg \, s^{-1} \, cm^{-2}}$, which is lower than the one used in this analysis. Since a lower flux limit also implies a lower cumulative bias, these values are underestimated and therefore we expect to find pessimistic constraints from the  artificially decreased term $b_A-1$. 

Moreover, since there is a large uncertainty on the properties of \oii\ galaxies, we also performed the analysis excluding these galaxies. In this case we discarded the four corresponding redshift bins not only for the galaxy auto-correlation but also from the \hi\ data set and in the multi-tracer analysis. This scenario should provide the most conservative results, since the multi-tracer likelihood also lacks the \hi\ single-tracer information from the non-overlapping redshift interval.

The results of this forecast are shown in the right panel of \cref{niter_bias} and are consistent with the previous tests: the galaxy auto-correlation is more prone to variations of the results while the multi-tracer is more stable. It is also interesting to note that a lower galaxy bias in the multi-tracer analysis impacts much less than cutting the redshift range at $z=2.6$, since a cut in redshift affects the full data vector and not only one of the two tracers. In this regard,  the different response to this cut in the single-tracer  and  multi-tracer cases is interesting. The \hi\ auto-correlation constraints respond the most, while the galaxy auto-correlation is less affected.  This is in line with the overall results of the analysis, where we find the strongest constraining power for \hi\ at the highest redshift, where galaxies are already limited by the shot noise (see \cref{constr_fgdampf2}). The loss of signal from this redshift range is much more damaging for the \hi\ single-tracer analysis. The impact on the multi-tracer method is, as expected, a balance between the one on the two auto-correlations. Furthermore, the variation of the galaxy bias has almost the same effects using the full $z$-range or using only the redshifts $z<2.6$.

In all the considered scenarios, the multi-tracer technique is confirmed to be the most powerful in constraining $\fnl$.

\section{Fisher matrix formalism}
{In analogy to the expression of the signal-to-noise ratio and of the likelihood defined in \cref{sec:snr,sec:analysis}, the Fisher matrix is 
\begin{equation} \label{fisher}
    \left( \mathcal{F}^{\rm MT}_{\rm overlap} \right)_{\rm \alpha\beta} = \left( \mathcal{F}^{\rm MT}_{\rm overlap} \right)_{\rm \alpha\beta} + \left( \mathcal{F}^{ AA}_{\rm no-overlap} \right)_{\rm \alpha\beta} + \left(\mathcal{F}^{BB}_{\rm no-overlap}\right)_{\rm \alpha\beta} \;.
\end{equation}
Here derivatives of the power spectrum are computed with respect to the parameter vector $ \theta_\alpha=\left\{ \fnl, n_{\rm s}, b_{\rm g}(z), b_{\rm \hi}(z)\right\}$. The multi-tracer overlap contribution and the single-tracer non-overlap contribution are 
\begin{align} \label{fisherMT}
    \left( \mathcal{F}^{\rm MT}_{\rm overlap} \right)_{\rm \alpha\beta} & = \sum_{k,\mu,z}  \partial_{\alpha}{\bm P}^{\sf T}\,\mathsf{Cov}^{-1}({\bm P},{\bm P})\,\partial_{\beta}{\bm P} \;,
\\ \label{fisherauto}
    \left( \mathcal{F}^{AB} \right)_{\rm \alpha\beta} &= \sum_{k,\mu,z}\frac{ \partial_{\alpha}P_{AB} \,\partial_{\beta}P_{AB}}{\left[\Delta P_{AB}\right]^2} \; .
\end{align}
The marginalised uncertainty on each free parameter of the model can be then evaluated as 
\begin{equation}   \sigma\left({\theta}_{\alpha}\right)=\sqrt{\Big( \mathcal{F}^{\rm MT}_{\rm overlap} \Big)_{\rm \alpha\alpha}^{-1}} \; ,
\end{equation}
while the conditional error is given by
\begin{equation}   \sigma\left({\theta}_{\alpha}\right)=\sqrt{\frac{1}{\Big( \mathcal{F}^{\rm MT}_{\rm overlap} \Big)_{\rm \alpha\alpha}}} \; .
\end{equation}


\providecommand{\href}[2]{#2}\begingroup\raggedright\begin{thebibliography}{10}

\bibitem{2014PDU.....5...75M}
J.~{Martin}, C.~{Ringeval} and V.~{Vennin}, \emph{{Encyclop{\ae}dia
  Inflationaris}},
  \href{https://doi.org/10.1016/j.dark.2014.01.003}{\emph{Physics of the Dark
  Universe} {\bfseries 5} (2014) 75}
  [\href{https://arxiv.org/abs/1303.3787}{{\ttfamily 1303.3787}}].

\bibitem{2022arXiv220308128A}
A.~{Ach{\'u}carro}, M.~{Biagetti}, M.~{Braglia}, G.~{Cabass}, R.~{Caldwell},
  E.~{Castorina} et~al., \emph{{Inflation: Theory and Observations}},
  \href{https://doi.org/10.48550/arXiv.2203.08128}{\emph{arXiv e-prints} (2022)
  arXiv:2203.08128} [\href{https://arxiv.org/abs/2203.08128}{{\ttfamily
  2203.08128}}].

\bibitem{2004PhR...402..103B}
N.~{Bartolo}, E.~{Komatsu}, S.~{Matarrese} and A.~{Riotto},
  \emph{{Non-Gaussianity from inflation: theory and observations}},
  \href{https://doi.org/10.1016/j.physrep.2004.08.022}{\emph{\physrep}
  {\bfseries 402} (2004) 103}
  [\href{https://arxiv.org/abs/astro-ph/0406398}{{\ttfamily
  astro-ph/0406398}}].

\bibitem{2018arXiv181208197C}
M.~{Celoria} and S.~{Matarrese}, \emph{{Primordial Non-Gaussianity}},
  \href{https://doi.org/10.48550/arXiv.1812.08197}{\emph{arXiv e-prints} (2018)
  arXiv:1812.08197} [\href{https://arxiv.org/abs/1812.08197}{{\ttfamily
  1812.08197}}].

\bibitem{2020A&A...641A...9P}
{Planck Collaboration}, Y.~{Akrami}, F.~{Arroja}, M.~{Ashdown}, J.~{Aumont},
  C.~{Baccigalupi} et~al., \emph{{Planck 2018 results. IX. Constraints on
  primordial non-Gaussianity}},
  \href{https://doi.org/10.1051/0004-6361/201935891}{\emph{\aap} {\bfseries
  641} (2020) A9} [\href{https://arxiv.org/abs/1905.05697}{{\ttfamily
  1905.05697}}].

\bibitem{2011JCAP...04..006B}
T.~{Baldauf}, U.~{Seljak} and L.~{Senatore}, \emph{{Primordial non-Gaussianity
  in the bispectrum of the halo density field}},
  \href{https://doi.org/10.1088/1475-7516/2011/04/006}{\emph{\jcap} {\bfseries
  2011} (2011) 006} [\href{https://arxiv.org/abs/1011.1513}{{\ttfamily
  1011.1513}}].

\bibitem{2011JCAP...11..038C}
P.~{Creminelli}, G.~{D'Amico}, M.~{Musso} and J.~{Nore{\~n}a}, \emph{{The (not
  so) squeezed limit of the primordial 3-point function}},
  \href{https://doi.org/10.1088/1475-7516/2011/11/038}{\emph{\jcap} {\bfseries
  2011} (2011) 038} [\href{https://arxiv.org/abs/1106.1462}{{\ttfamily
  1106.1462}}].

\bibitem{2013PhRvD..88h3502P}
E.~{Pajer}, F.~{Schmidt} and M.~{Zaldarriaga}, \emph{{The Observed squeezed
  limit of cosmological three-point functions}},
  \href{https://doi.org/10.1103/PhysRevD.88.083502}{\emph{\prd} {\bfseries 88}
  (2013) 083502} [\href{https://arxiv.org/abs/1305.0824}{{\ttfamily
  1305.0824}}].

\bibitem{2001PhRvD..63f3002K}
E.~{Komatsu} and D.N.~{Spergel}, \emph{{Acoustic signatures in the primary
  microwave background bispectrum}},
  \href{https://doi.org/10.1103/PhysRevD.63.063002}{\emph{\prd} {\bfseries 63}
  (2001) 063002} [\href{https://arxiv.org/abs/astro-ph/0005036}{{\ttfamily
  astro-ph/0005036}}].

\bibitem{2008PhRvD..77l3514D}
N.~{Dalal}, O.~{Dor{\'e}}, D.~{Huterer} and A.~{Shirokov}, \emph{{Imprints of
  primordial non-Gaussianities on large-scale structure: Scale-dependent bias
  and abundance of virialized objects}},
  \href{https://doi.org/10.1103/PhysRevD.77.123514}{\emph{\prd} {\bfseries 77}
  (2008) 123514} [\href{https://arxiv.org/abs/0710.4560}{{\ttfamily
  0710.4560}}].

\bibitem{2008ApJ...677L..77M}
S.~{Matarrese} and L.~{Verde}, \emph{{The Effect of Primordial Non-Gaussianity
  on Halo Bias}}, \href{https://doi.org/10.1086/587840}{\emph{\apjl} {\bfseries
  677} (2008) L77} [\href{https://arxiv.org/abs/0801.4826}{{\ttfamily
  0801.4826}}].

\bibitem{2015PhRvD..92f3525A}
D.~{Alonso} and P.G.~{Ferreira}, \emph{{Constraining ultralarge-scale cosmology
  with multiple tracers in optical and radio surveys}},
  \href{https://doi.org/10.1103/PhysRevD.92.063525}{\emph{\prd} {\bfseries 92}
  (2015) 063525} [\href{https://arxiv.org/abs/1507.03550}{{\ttfamily
  1507.03550}}].

\bibitem{2014MNRAS.442.2511F}
L.D.~{Ferramacho}, M.G.~{Santos}, M.J.~{Jarvis} and S.~{Camera}, \emph{{Radio
  galaxy populations and the multitracer technique: pushing the limits on
  primordial non-Gaussianity}},
  \href{https://doi.org/10.1093/mnras/stu1015}{\emph{\mnras} {\bfseries 442}
  (2014) 2511} [\href{https://arxiv.org/abs/1402.2290}{{\ttfamily 1402.2290}}].

\bibitem{2021JCAP...06..039J}
S.~{Jolicoeur}, R.~{Maartens}, E.M.~{De Weerd}, O.~{Umeh}, C.~{Clarkson} and
  S.~{Camera}, \emph{{Detecting the relativistic bispectrum in 21cm intensity
  maps}}, \href{https://doi.org/10.1088/1475-7516/2021/06/039}{\emph{\jcap}
  {\bfseries 2021} (2021) 039}
  [\href{https://arxiv.org/abs/2009.06197}{{\ttfamily 2009.06197}}].

\bibitem{2015ApJ...812L..22F}
J.~{Fonseca}, S.~{Camera}, M.G.~{Santos} and R.~{Maartens}, \emph{{Hunting Down
  Horizon-scale Effects with Multi-wavelength Surveys}},
  \href{https://doi.org/10.1088/2041-8205/812/2/L22}{\emph{\apjl} {\bfseries
  812} (2015) L22} [\href{https://arxiv.org/abs/1507.04605}{{\ttfamily
  1507.04605}}].

\bibitem{2020MNRAS.492.1513G}
Z.~{Gomes}, S.~{Camera}, M.J.~{Jarvis}, C.~{Hale} and J.~{Fonseca},
  \emph{{Non-Gaussianity constraints using future radio continuum surveys and
  the multitracer technique}},
  \href{https://doi.org/10.1093/mnras/stz3581}{\emph{\mnras} {\bfseries 492}
  (2020) 1513} [\href{https://arxiv.org/abs/1912.08362}{{\ttfamily
  1912.08362}}].

\bibitem{2021JCAP...11..010V}
J.-A.~{Viljoen}, J.~{Fonseca} and R.~{Maartens}, \emph{{Multi-wavelength
  spectroscopic probes: prospects for primordial non-Gaussianity and
  relativistic effects}},
  \href{https://doi.org/10.1088/1475-7516/2021/11/010}{\emph{\jcap} {\bfseries
  2021} (2021) 010} [\href{https://arxiv.org/abs/2107.14057}{{\ttfamily
  2107.14057}}].

\bibitem{2020JCAP...12..031B}
A.~{Barreira}, \emph{{On the impact of galaxy bias uncertainties on primordial
  non-Gaussianity constraints}},
  \href{https://doi.org/10.1088/1475-7516/2020/12/031}{\emph{\jcap} {\bfseries
  2020} (2020) 031} [\href{https://arxiv.org/abs/2009.06622}{{\ttfamily
  2009.06622}}].

\bibitem{2022JCAP...01..033B}
A.~{Barreira}, \emph{{Predictions for local PNG bias in the galaxy power
  spectrum and bispectrum and the consequences for f $_{NL}$ constraints}},
  \href{https://doi.org/10.1088/1475-7516/2022/01/033}{\emph{\jcap} {\bfseries
  2022} (2022) 033} [\href{https://arxiv.org/abs/2107.06887}{{\ttfamily
  2107.06887}}].

\bibitem{2023arXiv230209066B}
A.~{Barreira} and E.~{Krause}, \emph{{Towards optimal and robust $f_{\rm NL}$
  constraints with multi-tracer analyses}},
  \href{https://doi.org/10.48550/arXiv.2302.09066}{\emph{arXiv e-prints} (2023)
  arXiv:2302.09066} [\href{https://arxiv.org/abs/2302.09066}{{\ttfamily
  2302.09066}}].

\bibitem{2018MNRAS.478.1341K}
D.~{Karagiannis}, A.~{Lazanu}, M.~{Liguori}, A.~{Raccanelli}, N.~{Bartolo} and
  L.~{Verde}, \emph{{Constraining primordial non-Gaussianity with bispectrum
  and power spectrum from upcoming optical and radio surveys}},
  \href{https://doi.org/10.1093/mnras/sty1029}{\emph{\mnras} {\bfseries 478}
  (2018) 1341} [\href{https://arxiv.org/abs/1801.09280}{{\ttfamily
  1801.09280}}].

\bibitem{2009PhRvL.102b1302S}
U.~{Seljak}, \emph{{Extracting Primordial Non-Gaussianity without Cosmic
  Variance}}, \href{https://doi.org/10.1103/PhysRevLett.102.021302}{\emph{\prl}
  {\bfseries 102} (2009) 021302}
  [\href{https://arxiv.org/abs/0807.1770}{{\ttfamily 0807.1770}}].

\bibitem{2013MNRAS.432..318A}
L.R.~{Abramo} and K.E.~{Leonard}, \emph{{Why multitracer surveys beat cosmic
  variance}}, \href{https://doi.org/10.1093/mnras/stt465}{\emph{\mnras}
  {\bfseries 432} (2013) 318}
  [\href{https://arxiv.org/abs/1302.5444}{{\ttfamily 1302.5444}}].

\bibitem{2009JCAP...10..007M}
P.~{McDonald} and U.~{Seljak}, \emph{{How to evade the sample variance limit on
  measurements of redshift-space distortions}},
  \href{https://doi.org/10.1088/1475-7516/2009/10/007}{\emph{\jcap} {\bfseries
  2009} (2009) 007} [\href{https://arxiv.org/abs/0810.0323}{{\ttfamily
  0810.0323}}].

\bibitem{2019MNRAS.489.1950B}
M.~{Ballardini}, W.L.~{Matthewson} and R.~{Maartens}, \emph{{Constraining
  primordial non-Gaussianity using two galaxy surveys and CMB lensing}},
  \href{https://doi.org/10.1093/mnras/stz2258}{\emph{\mnras} {\bfseries 489}
  (2019) 1950} [\href{https://arxiv.org/abs/1906.04730}{{\ttfamily
  1906.04730}}].

\bibitem{2008PhRvD..78l3519M}
P.~{McDonald}, \emph{{Primordial non-Gaussianity: Large-scale structure
  signature in the perturbative bias model}},
  \href{https://doi.org/10.1103/PhysRevD.78.123519}{\emph{\prd} {\bfseries 78}
  (2008) 123519} [\href{https://arxiv.org/abs/0806.1061}{{\ttfamily
  0806.1061}}].

\bibitem{2021JCAP...05..015M}
A.~{Moradinezhad Dizgah}, M.~{Biagetti}, E.~{Sefusatti}, V.~{Desjacques} and
  J.~{Nore{\~n}a}, \emph{{Primordial non-Gaussianity from biased tracers:
  likelihood analysis of real-space power spectrum and bispectrum}},
  \href{https://doi.org/10.1088/1475-7516/2021/05/015}{\emph{\jcap} {\bfseries
  2021} (2021) 015} [\href{https://arxiv.org/abs/2010.14523}{{\ttfamily
  2010.14523}}].

\bibitem{2023JCAP...01..023L}
T.~{Lazeyras}, A.~{Barreira}, F.~{Schmidt} and V.~{Desjacques}, \emph{{Assembly
  bias in the local PNG halo bias and its implication for f $_{NL}$
  constraints}},
  \href{https://doi.org/10.1088/1475-7516/2023/01/023}{\emph{\jcap} {\bfseries
  2023} (2023) 023} [\href{https://arxiv.org/abs/2209.07251}{{\ttfamily
  2209.07251}}].

\bibitem{2022PhRvD.106d3506C}
G.~{Cabass}, M.M.~{Ivanov}, O.H.E.~{Philcox}, M.~{Simonovi{\'c}} and
  M.~{Zaldarriaga}, \emph{{Constraints on multifield inflation from the BOSS
  galaxy survey}},
  \href{https://doi.org/10.1103/PhysRevD.106.043506}{\emph{\prd} {\bfseries
  106} (2022) 043506} [\href{https://arxiv.org/abs/2204.01781}{{\ttfamily
  2204.01781}}].

\bibitem{2022JCAP...11..013B}
A.~{Barreira}, \emph{{Can we actually constrain f$_{NL}$ using the
  scale-dependent bias effect? An illustration of the impact of galaxy bias
  uncertainties using the BOSS DR12 galaxy power spectrum}},
  \href{https://doi.org/10.1088/1475-7516/2022/11/013}{\emph{\jcap} {\bfseries
  2022} (2022) 013} [\href{https://arxiv.org/abs/2205.05673}{{\ttfamily
  2205.05673}}].

\bibitem{2023arXiv230504028K}
D.~{Karagiannis}, R.~{Maartens}, J.~{Fonseca}, S.~{Camera} and C.~{Clarkson},
  \emph{{Multi-tracer power spectra and bispectra: Formalism}},
  \href{https://doi.org/10.48550/arXiv.2305.04028}{\emph{arXiv e-prints} (2023)
  arXiv:2305.04028} [\href{https://arxiv.org/abs/2305.04028}{{\ttfamily
  2305.04028}}].

\bibitem{2010PhRvD..82h3508Y}
J.~{Yoo}, \emph{{General relativistic description of the observed galaxy power
  spectrum: Do we understand what we measure?}},
  \href{https://doi.org/10.1103/PhysRevD.82.083508}{\emph{\prd} {\bfseries 82}
  (2010) 083508} [\href{https://arxiv.org/abs/1009.3021}{{\ttfamily
  1009.3021}}].

\bibitem{2011PhRvD..84f3505B}
C.~{Bonvin} and R.~{Durrer}, \emph{{What galaxy surveys really measure}},
  \href{https://doi.org/10.1103/PhysRevD.84.063505}{\emph{\prd} {\bfseries 84}
  (2011) 063505} [\href{https://arxiv.org/abs/1105.5280}{{\ttfamily
  1105.5280}}].

\bibitem{2011PhRvD..84d3516C}
A.~{Challinor} and A.~{Lewis}, \emph{{Linear power spectrum of observed source
  number counts}},
  \href{https://doi.org/10.1103/PhysRevD.84.043516}{\emph{\prd} {\bfseries 84}
  (2011) 043516} [\href{https://arxiv.org/abs/1105.5292}{{\ttfamily
  1105.5292}}].

\bibitem{2012PhRvD..85d1301B}
M.~{Bruni}, R.~{Crittenden}, K.~{Koyama}, R.~{Maartens}, C.~{Pitrou} and
  D.~{Wands}, \emph{{Disentangling non-Gaussianity, bias, and general
  relativistic effects in the galaxy distribution}},
  \href{https://doi.org/10.1103/PhysRevD.85.041301}{\emph{\prd} {\bfseries 85}
  (2012) 041301} [\href{https://arxiv.org/abs/1106.3999}{{\ttfamily
  1106.3999}}].

\bibitem{2012PhRvD..85b3504J}
D.~{Jeong}, F.~{Schmidt} and C.M.~{Hirata}, \emph{{Large-scale clustering of
  galaxies in general relativity}},
  \href{https://doi.org/10.1103/PhysRevD.85.023504}{\emph{\prd} {\bfseries 85}
  (2012) 023504} [\href{https://arxiv.org/abs/1107.5427}{{\ttfamily
  1107.5427}}].

\bibitem{2012PhRvD..86f3514Y}
J.~{Yoo}, N.~{Hamaus}, U.~{Seljak} and M.~{Zaldarriaga}, \emph{{Going beyond
  the Kaiser redshift-space distortion formula: A full general relativistic
  account of the effects and their detectability in galaxy clustering}},
  \href{https://doi.org/10.1103/PhysRevD.86.063514}{\emph{\prd} {\bfseries 86}
  (2012) 063514} [\href{https://arxiv.org/abs/1109.0998}{{\ttfamily
  1109.0998}}].

\bibitem{2015MNRAS.448.1035C}
S.~{Camera}, M.G.~{Santos} and R.~{Maartens}, \emph{{Probing primordial
  non-Gaussianity with SKA galaxy redshift surveys: a fully relativistic
  analysis}}, \href{https://doi.org/10.1093/mnras/stv040}{\emph{\mnras}
  {\bfseries 448} (2015) 1035}
  [\href{https://arxiv.org/abs/1409.8286}{{\ttfamily 1409.8286}}].

\bibitem{2015ApJ...814..145A}
D.~{Alonso}, P.~{Bull}, P.G.~{Ferreira}, R.~{Maartens} and M.G.~{Santos},
  \emph{{Ultra-large-scale Cosmology in Next-generation Experiments with Single
  Tracers}}, \href{https://doi.org/10.1088/0004-637X/814/2/145}{\emph{\apj}
  {\bfseries 814} (2015) 145}
  [\href{https://arxiv.org/abs/1505.07596}{{\ttfamily 1505.07596}}].

\bibitem{2011PhRvD..83l3514N}
T.~{Namikawa}, T.~{Okamura} and A.~{Taruya}, \emph{{Magnification effect on the
  detection of primordial non-Gaussianity from photometric surveys}},
  \href{https://doi.org/10.1103/PhysRevD.83.123514}{\emph{\prd} {\bfseries 83}
  (2011) 123514} [\href{https://arxiv.org/abs/1103.1118}{{\ttfamily
  1103.1118}}].

\bibitem{2022A&A...662A..93E}
{Euclid Collaboration}, F.~{Lepori}, I.~{Tutusaus}, C.~{Viglione}, C.~{Bonvin},
  S.~{Camera} et~al., \emph{{Euclid preparation. XIX. Impact of magnification
  on photometric galaxy clustering}},
  \href{https://doi.org/10.1051/0004-6361/202142419}{\emph{\aap} {\bfseries
  662} (2022) A93} [\href{https://arxiv.org/abs/2110.05435}{{\ttfamily
  2110.05435}}].

\bibitem{2015MNRAS.451L..80C}
S.~{Camera}, R.~{Maartens} and M.G.~{Santos}, \emph{{Einstein's legacy in
  galaxy surveys.}}, \href{https://doi.org/10.1093/mnrasl/slv069}{\emph{\mnras}
  {\bfseries 451} (2015) L80}
  [\href{https://arxiv.org/abs/1412.4781}{{\ttfamily 1412.4781}}].

\bibitem{2021JCAP...12..004V}
J.-A.~{Viljoen}, J.~{Fonseca} and R.~{Maartens}, \emph{{Multi-wavelength
  spectroscopic probes: biases from neglecting light-cone effects}},
  \href{https://doi.org/10.1088/1475-7516/2021/12/004}{\emph{\jcap} {\bfseries
  2021} (2021) 004} [\href{https://arxiv.org/abs/2108.05746}{{\ttfamily
  2108.05746}}].

\bibitem{2023PhRvL.131k1201F}
M.~{Foglieni}, M.~{Pantiri}, E.~{Di Dio} and E.~{Castorina}, \emph{{Large Scale
  Limit of the Observed Galaxy Power Spectrum}},
  \href{https://doi.org/10.1103/PhysRevLett.131.111201}{\emph{\prl} {\bfseries
  131} (2023) 111201} [\href{https://arxiv.org/abs/2303.03142}{{\ttfamily
  2303.03142}}].

\bibitem{2003MNRAS.341.1311S}
R.E.~{Smith}, J.A.~{Peacock}, A.~{Jenkins}, S.D.M.~{White}, C.S.~{Frenk},
  F.R.~{Pearce} et~al., \emph{{Stable clustering, the halo model and non-linear
  cosmological power spectra}},
  \href{https://doi.org/10.1046/j.1365-8711.2003.06503.x}{\emph{\mnras}
  {\bfseries 341} (2003) 1311}
  [\href{https://arxiv.org/abs/astro-ph/0207664}{{\ttfamily
  astro-ph/0207664}}].

\bibitem{2020JCAP...03..065M}
R.~{Maartens}, S.~{Jolicoeur}, O.~{Umeh}, E.M.~{De Weerd}, C.~{Clarkson} and
  S.~{Camera}, \emph{{Detecting the relativistic galaxy bispectrum}},
  \href{https://doi.org/10.1088/1475-7516/2020/03/065}{\emph{\jcap} {\bfseries
  2020} (2020) 065} [\href{https://arxiv.org/abs/1911.02398}{{\ttfamily
  1911.02398}}].

\bibitem{2020MNRAS.495.1340F}
J.~{Fonseca} and S.~{Camera}, \emph{{High-redshift cosmology with oxygen lines
  from H{\ensuremath{\alpha}} surveys}},
  \href{https://doi.org/10.1093/mnras/staa1136}{\emph{\mnras} {\bfseries 495}
  (2020) 1340} [\href{https://arxiv.org/abs/2001.04473}{{\ttfamily
  2001.04473}}].

\bibitem{2021MNRAS.505.2784Z}
Z.~{Zhai}, Y.~{Wang}, A.~{Benson}, C.-H.~{Chuang} and G.~{Yepes}, \emph{{Linear
  bias and halo occupation distribution of emission-line galaxies from Nancy
  Grace Roman Space Telescope}},
  \href{https://doi.org/10.1093/mnras/stab1539}{\emph{\mnras} {\bfseries 505}
  (2021) 2784} [\href{https://arxiv.org/abs/2103.11063}{{\ttfamily
  2103.11063}}].

\bibitem{2023MNRAS.523.2498M}
K.S.~{McCarthy}, Z.~{Zhai} and Y.~{Wang}, \emph{{Phenomenological power
  spectrum models for H {\ensuremath{\alpha}} emission line galaxies from the
  Nancy Grace Roman Space Telescope}},
  \href{https://doi.org/10.1093/mnras/stad1591}{\emph{\mnras} {\bfseries 523}
  (2023) 2498} [\href{https://arxiv.org/abs/2212.08699}{{\ttfamily
  2212.08699}}].

\bibitem{2020MNRAS.493..747P}
H.~{Pan}, D.~{Obreschkow}, C.~{Howlett}, C.d.P.~{Lagos}, P.J.~{Elahi},
  C.~{Baugh} et~al., \emph{{Multiwavelength consensus of large-scale linear
  bias}}, \href{https://doi.org/10.1093/mnras/staa222}{\emph{\mnras} {\bfseries
  493} (2020) 747} [\href{https://arxiv.org/abs/1909.12069}{{\ttfamily
  1909.12069}}].

\bibitem{2016A&A...590A...3P}
L.~{Pozzetti}, C.M.~{Hirata}, J.E.~{Geach}, A.~{Cimatti}, C.~{Baugh},
  O.~{Cucciati} et~al., \emph{{Modelling the number density of
  H{\ensuremath{\alpha}} emitters for future spectroscopic near-IR space
  missions}}, \href{https://doi.org/10.1051/0004-6361/201527081}{\emph{\aap}
  {\bfseries 590} (2016) A3}
  [\href{https://arxiv.org/abs/1603.01453}{{\ttfamily 1603.01453}}].

\bibitem{2015MNRAS.452.3948K}
A.A.~{Khostovan}, D.~{Sobral}, B.~{Mobasher}, P.N.~{Best}, I.~{Smail},
  J.P.~{Stott} et~al., \emph{{Evolution of the H {\ensuremath{\beta}} + [O III]
  and [O II] luminosity functions and the [O II] star formation history of the
  Universe up to z {\ensuremath{\sim}} 5 from HiZELS}},
  \href{https://doi.org/10.1093/mnras/stv1474}{\emph{\mnras} {\bfseries 452}
  (2015) 3948} [\href{https://arxiv.org/abs/1503.00004}{{\ttfamily
  1503.00004}}].

\bibitem{2011arXiv1110.3193L}
R.~{Laureijs}, J.~{Amiaux}, S.~{Arduini}, J.L.~{Augu{\`e}res}, J.~{Brinchmann},
  R.~{Cole} et~al., \emph{{Euclid Definition Study Report}},
  \href{https://doi.org/10.48550/arXiv.1110.3193}{\emph{arXiv e-prints} (2011)
  arXiv:1110.3193} [\href{https://arxiv.org/abs/1110.3193}{{\ttfamily
  1110.3193}}].

\bibitem{2022ApJ...928....1W}
Y.~{Wang}, Z.~{Zhai}, A.~{Alavi}, E.~{Massara}, A.~{Pisani}, A.~{Benson}
  et~al., \emph{{The High Latitude Spectroscopic Survey on the Nancy Grace
  Roman Space Telescope}},
  \href{https://doi.org/10.3847/1538-4357/ac4973}{\emph{\apj} {\bfseries 928}
  (2022) 1} [\href{https://arxiv.org/abs/2110.01829}{{\ttfamily 2110.01829}}].

\bibitem{2020PASA...37....7S}
{Square Kilometre Array Cosmology Science Working Group}, D.J.~{Bacon},
  R.A.~{Battye}, P.~{Bull}, S.~{Camera}, P.G.~{Ferreira} et~al.,
  \emph{{Cosmology with Phase 1 of the Square Kilometre Array Red Book 2018:
  Technical specifications and performance forecasts}},
  \href{https://doi.org/10.1017/pasa.2019.51}{\emph{\pasa} {\bfseries 37}
  (2020) e007} [\href{https://arxiv.org/abs/1811.02743}{{\ttfamily
  1811.02743}}].

\bibitem{2018ApJ...866..135V}
F.~{Villaescusa-Navarro}, S.~{Genel}, E.~{Castorina}, A.~{Obuljen},
  D.N.~{Spergel}, L.~{Hernquist} et~al., \emph{{Ingredients for 21 cm Intensity
  Mapping}}, \href{https://doi.org/10.3847/1538-4357/aadba0}{\emph{\apj}
  {\bfseries 866} (2018) 135}
  [\href{https://arxiv.org/abs/1804.09180}{{\ttfamily 1804.09180}}].

\bibitem{2016mks..confE..32S}
M.~{Santos}, P.~{Bull}, S.~{Camera}, S.~{Chen}, J.~{Fonseca}, I.~{Heywood}
  et~al., \emph{{A Large Sky Survey with MeerKAT}},  in \emph{MeerKAT Science:
  On the Pathway to the SKA}, p.~32, Jan., 2016,
  \href{https://doi.org/10.22323/1.277.0032}{DOI}
  [\href{https://arxiv.org/abs/1709.06099}{{\ttfamily 1709.06099}}].

\bibitem{2019JCAP...12..028F}
J.~{Fonseca}, J.-A.~{Viljoen} and R.~{Maartens}, \emph{{Constraints on the
  growth rate using the observed galaxy power spectrum}},
  \href{https://doi.org/10.1088/1475-7516/2019/12/028}{\emph{\jcap} {\bfseries
  2019} (2019) 028} [\href{https://arxiv.org/abs/1907.02975}{{\ttfamily
  1907.02975}}].

\bibitem{2015aska.confE..19S}
M.~{Santos}, P.~{Bull}, D.~{Alonso}, S.~{Camera}, P.~{Ferreira}, G.~{Bernardi}
  et~al., \emph{{Cosmology from a SKA HI intensity mapping survey}},  in
  \emph{Advancing Astrophysics with the Square Kilometre Array (AASKA14)},
  p.~19, Apr., 2015, \href{https://doi.org/10.22323/1.215.0019}{DOI}
  [\href{https://arxiv.org/abs/1501.03989}{{\ttfamily 1501.03989}}].

\bibitem{2017MNRAS.471.1788C}
E.~{Castorina} and F.~{Villaescusa-Navarro}, \emph{{On the spatial distribution
  of neutral hydrogen in the Universe: bias and shot-noise of the H I power
  spectrum}}, \href{https://doi.org/10.1093/mnras/stx1599}{\emph{\mnras}
  {\bfseries 471} (2017) 1788}
  [\href{https://arxiv.org/abs/1609.05157}{{\ttfamily 1609.05157}}].

\bibitem{2023MNRAS.518.6262C}
S.~{Cunnington}, Y.~{Li}, M.G.~{Santos}, J.~{Wang}, I.P.~{Carucci},
  M.O.~{Irfan} et~al., \emph{{H I intensity mapping with MeerKAT: power
  spectrum detection in cross-correlation with WiggleZ galaxies}},
  \href{https://doi.org/10.1093/mnras/stac3060}{\emph{\mnras} {\bfseries 518}
  (2023) 6262} [\href{https://arxiv.org/abs/2206.01579}{{\ttfamily
  2206.01579}}].

\bibitem{2023MNRAS.523.2453C}
S.~{Cunnington}, L.~{Wolz}, P.~{Bull}, I.P.~{Carucci}, K.~{Grainge},
  M.O.~{Irfan} et~al., \emph{{The foreground transfer function for H I
  intensity mapping signal reconstruction: MeerKLASS and precision cosmology
  applications}}, \href{https://doi.org/10.1093/mnras/stad1567}{\emph{\mnras}
  {\bfseries 523} (2023) 2453}
  [\href{https://arxiv.org/abs/2302.07034}{{\ttfamily 2302.07034}}].

\bibitem{2019MNRAS.489..153B}
C.~{Blake}, \emph{{Power spectrum modelling of galaxy and radio intensity maps
  including observational effects}},
  \href{https://doi.org/10.1093/mnras/stz2145}{\emph{\mnras} {\bfseries 489}
  (2019) 153} [\href{https://arxiv.org/abs/1902.07439}{{\ttfamily
  1902.07439}}].

\bibitem{2015ApJ...803...21B}
P.~{Bull}, P.G.~{Ferreira}, P.~{Patel} and M.G.~{Santos}, \emph{{Late-time
  Cosmology with 21 cm Intensity Mapping Experiments}},
  \href{https://doi.org/10.1088/0004-637X/803/1/21}{\emph{\apj} {\bfseries 803}
  (2015) 21} [\href{https://arxiv.org/abs/1405.1452}{{\ttfamily 1405.1452}}].

\bibitem{2014MNRAS.441.3271W}
L.~{Wolz}, F.B.~{Abdalla}, C.~{Blake}, J.R.~{Shaw}, E.~{Chapman} and
  S.~{Rawlings}, \emph{{The effect of foreground subtraction on cosmological
  measurements from intensity mapping}},
  \href{https://doi.org/10.1093/mnras/stu792}{\emph{\mnras} {\bfseries 441}
  (2014) 3271} [\href{https://arxiv.org/abs/1310.8144}{{\ttfamily 1310.8144}}].

\bibitem{2019MNRAS.488.5452C}
S.~{Cunnington}, L.~{Wolz}, A.~{Pourtsidou} and D.~{Bacon}, \emph{{Impact of
  foregrounds on H I intensity mapping cross-correlations with optical
  surveys}}, \href{https://doi.org/10.1093/mnras/stz1916}{\emph{\mnras}
  {\bfseries 488} (2019) 5452}
  [\href{https://arxiv.org/abs/1904.01479}{{\ttfamily 1904.01479}}].

\bibitem{2020MNRAS.499.4054C}
S.~{Cunnington}, S.~{Camera} and A.~{Pourtsidou}, \emph{{The degeneracy between
  primordial non-Gaussianity and foregrounds in 21 cm intensity mapping
  experiments}}, \href{https://doi.org/10.1093/mnras/staa2986}{\emph{\mnras}
  {\bfseries 499} (2020) 4054}
  [\href{https://arxiv.org/abs/2007.12126}{{\ttfamily 2007.12126}}].

\bibitem{2022MNRAS.509.2048S}
M.~{Spinelli}, I.P.~{Carucci}, S.~{Cunnington}, S.E.~{Harper}, M.O.~{Irfan},
  J.~{Fonseca} et~al., \emph{{SKAO H I intensity mapping: blind foreground
  subtraction challenge}},
  \href{https://doi.org/10.1093/mnras/stab3064}{\emph{\mnras} {\bfseries 509}
  (2022) 2048} [\href{https://arxiv.org/abs/2107.10814}{{\ttfamily
  2107.10814}}].

\bibitem{2022MNRAS.512.2408C}
S.~{Cunnington}, \emph{{Detecting the power spectrum turnover with H I
  intensity mapping}},
  \href{https://doi.org/10.1093/mnras/stac576}{\emph{\mnras} {\bfseries 512}
  (2022) 2408} [\href{https://arxiv.org/abs/2202.13828}{{\ttfamily
  2202.13828}}].

\bibitem{2021MNRAS.502.2549S}
P.S.~{Soares}, S.~{Cunnington}, A.~{Pourtsidou} and C.~{Blake}, \emph{{Power
  spectrum multipole expansion for H I intensity mapping experiments: unbiased
  parameter estimation}},
  \href{https://doi.org/10.1093/mnras/stab027}{\emph{\mnras} {\bfseries 502}
  (2021) 2549} [\href{https://arxiv.org/abs/2008.12102}{{\ttfamily
  2008.12102}}].

\bibitem{2023EPJC...83..320J}
S.~{Jolicoeur}, R.~{Maartens} and S.~{Dlamini}, \emph{{Constraining primordial
  non-Gaussianity by combining next-generation galaxy and 21 cm intensity
  mapping surveys}},
  \href{https://doi.org/10.1140/epjc/s10052-023-11482-2}{\emph{European
  Physical Journal C} {\bfseries 83} (2023) 320}
  [\href{https://arxiv.org/abs/2301.02406}{{\ttfamily 2301.02406}}].

\bibitem{2020JCAP...09..054V}
J.-A.~{Viljoen}, J.~{Fonseca} and R.~{Maartens}, \emph{{Constraining the growth
  rate by combining multiple future surveys}},
  \href{https://doi.org/10.1088/1475-7516/2020/09/054}{\emph{\jcap} {\bfseries
  2020} (2020) 054} [\href{https://arxiv.org/abs/2007.04656}{{\ttfamily
  2007.04656}}].

\bibitem{2013PASP..125..306F}
D.~{Foreman-Mackey}, D.W.~{Hogg}, D.~{Lang} and J.~{Goodman}, \emph{{emcee: The
  MCMC Hammer}}, \href{https://doi.org/10.1086/670067}{\emph{\pasp} {\bfseries
  125} (2013) 306} [\href{https://arxiv.org/abs/1202.3665}{{\ttfamily
  1202.3665}}].

\bibitem{2020A&A...641A...6P}
{Planck Collaboration}, N.~{Aghanim}, Y.~{Akrami}, M.~{Ashdown}, J.~{Aumont},
  C.~{Baccigalupi} et~al., \emph{{Planck 2018 results. VI. Cosmological
  parameters}}, \href{https://doi.org/10.1051/0004-6361/201833910}{\emph{\aap}
  {\bfseries 641} (2020) A6}
  [\href{https://arxiv.org/abs/1807.06209}{{\ttfamily 1807.06209}}].

\bibitem{2017PhRvD..95l3507D}
R.~{de Putter}, J.~{Gleyzes} and O.~{Dor{\'e}}, \emph{{Next non-Gaussianity
  frontier: What can a measurement with {\ensuremath{\sigma}}
  (f$_{NL}$){\ensuremath{\lesssim}}1 tell us about multifield inflation?}},
  \href{https://doi.org/10.1103/PhysRevD.95.123507}{\emph{\prd} {\bfseries 95}
  (2017) 123507} [\href{https://arxiv.org/abs/1612.05248}{{\ttfamily
  1612.05248}}].

\bibitem{2014arXiv1412.4671A}
M.~{Alvarez}, T.~{Baldauf}, J.R.~{Bond}, N.~{Dalal}, R.~{de Putter},
  O.~{Dor{\'e}} et~al., \emph{{Testing Inflation with Large Scale Structure:
  Connecting Hopes with Reality}},
  \href{https://doi.org/10.48550/arXiv.1412.4671}{\emph{arXiv e-prints} (2014)
  arXiv:1412.4671} [\href{https://arxiv.org/abs/1412.4671}{{\ttfamily
  1412.4671}}].

\bibitem{2020JCAP...12..013B}
A.~{Barreira}, G.~{Cabass}, F.~{Schmidt}, A.~{Pillepich} and D.~{Nelson},
  \emph{{Galaxy bias and primordial non-Gaussianity: insights from galaxy
  formation simulations with IllustrisTNG}},
  \href{https://doi.org/10.1088/1475-7516/2020/12/013}{\emph{\jcap} {\bfseries
  2020} (2020) 013} [\href{https://arxiv.org/abs/2006.09368}{{\ttfamily
  2006.09368}}].

\bibitem{2020MNRAS.499..304C}
I.P.~{Carucci}, M.O.~{Irfan} and J.~{Bobin}, \emph{{Recovery of 21-cm intensity
  maps with sparse component separation}},
  \href{https://doi.org/10.1093/mnras/staa2854}{\emph{\mnras} {\bfseries 499}
  (2020) 304} [\href{https://arxiv.org/abs/2006.05996}{{\ttfamily
  2006.05996}}].

\bibitem{2021MNRAS.504..208C}
S.~{Cunnington}, M.O.~{Irfan}, I.P.~{Carucci}, A.~{Pourtsidou} and J.~{Bobin},
  \emph{{21-cm foregrounds and polarization leakage: cleaning and mitigation
  strategies}}, \href{https://doi.org/10.1093/mnras/stab856}{\emph{\mnras}
  {\bfseries 504} (2021) 208}
  [\href{https://arxiv.org/abs/2010.02907}{{\ttfamily 2010.02907}}].

\bibitem{2014MNRAS.443..799O}
{\'A}.~{Orsi}, N.~{Padilla}, B.~{Groves}, S.~{Cora}, T.~{Tecce}, I.~{Gargiulo}
  et~al., \emph{{The nebular emission of star-forming galaxies in a
  hierarchical universe}},
  \href{https://doi.org/10.1093/mnras/stu1203}{\emph{\mnras} {\bfseries 443}
  (2014) 799} [\href{https://arxiv.org/abs/1402.5145}{{\ttfamily 1402.5145}}].

\end{thebibliography}\endgroup

\providecommand{\href}[2]{#2}\begingroup\raggedright\endgroup

\end{document}